\documentclass[twocolumn,twocolappendix]{aastex631}

\usepackage[utf8]{inputenc}
\usepackage{amsmath, amsfonts, amssymb, graphics,graphicx, wrapfig, nicefrac}
\usepackage{epsfig}
\usepackage{epstopdf}
\usepackage{multirow}
\usepackage{graphicx}
\usepackage{rotating}
\usepackage{mathrsfs}

\shorttitle{The CRIR in Diffuse Clouds}
\shortauthors{Luo et al.}
\graphicspath{{./}{figures/}}

\begin{document}

\title{Dependence of Molecular Abundance Ratios on the Cosmic Ray Ionization Rate in Nearby Diffuse Clouds}
\title{Molecular Abundance Ratios as a Probe of Cosmic Ray Ionization Rate in Diffuse Clouds}
\title{Abundance ratios of OH/CO and HCO$^+$/CO as probes of the cosmic ray ionization rate in diffuse clouds}

\correspondingauthor{Thomas G. Bisbas, Gan Luo}
\email{tbisbas@zhejianglab.com, luogan@nju.edu.cn}

\author[0000-0002-1583-8514]{Gan Luo}
\affiliation{School of Astronomy and Space Science, Nanjing University, Nanjing 210093, People’s Republic of China}
\affiliation{Key Laboratory of Modern Astronomy and Astrophysics (Nanjing University), Ministry of Education, Nanjing 210093, People’s Republic of China}

\author[0000-0002-7299-2876]{Zhi-Yu Zhang}
\affiliation{School of Astronomy and Space Science, Nanjing University, Nanjing 210093, People’s Republic of China}
\affiliation{Key Laboratory of Modern Astronomy and Astrophysics (Nanjing University), Ministry of Education, Nanjing 210093, People’s Republic of China}

\author[0000-0003-2733-4580]{Thomas G. Bisbas}
\affiliation{Research Center for Intelligent Computing Platforms, Zhejiang Laboratory, Hangzhou 311100, China}

\author[0000-0003-3010-7661]{Di Li}
\affiliation{CAS Key Laboratory of FAST, National Astronomical Observatories, Chinese Academy of Sciences, Beijing 100101, China}
\affiliation{Research Center for Intelligent Computing Platforms, Zhejiang Laboratory, Hangzhou 311100, China}
\affiliation{NAOC-UKZN Computational Astrophysics Centre, University of KwaZulu-Natal, Durban 4000, South Africa}

\author[0000-0002-5683-822X]{Ping Zhou}
\affiliation{School of Astronomy and Space Science, Nanjing University, Nanjing 210093, People’s Republic of China}
\affiliation{Key Laboratory of Modern Astronomy and Astrophysics (Nanjing University), Ministry of Education, Nanjing 210093, People’s Republic of China}

\author[0000-0002-2169-0472]{Ningyu Tang}
\affiliation{Department of Physics, Anhui Normal University, Wuhu, Anhui 241002, China}

\author[0000-0001-6106-1171]{Junzhi Wang}
\affiliation{School of Physical Science and Technology, Guangxi University, Nanning 530004, People's Republic of China}

\author[0000-0003-3948-9192]{Pei Zuo}
\affiliation{Kavli Institute for Astronomy and Astrophysics, Peking University, Beijing, 5 Yiheyuan Road, Haidian District, Beijing 100871, China}
\affiliation{International Centre for Radio Astronomy Research (ICRAR), University of Western Australia, Crawley, WA 6009, Australia}

\author[0000-0003-0355-6875]{Nannan Yue}
\affiliation{Kavli Institute for Astronomy and Astrophysics, Peking University, Beijing, 5 Yiheyuan Road, Haidian District, Beijing 100871, China}





\begin{abstract}

The cosmic-ray ionization rate (CRIR, $\zeta_2$) is one of the key parameters { controlling} the formation and destruction of various molecules in molecular clouds. However, the current most commonly used { CRIR tracers}, such as H$_3^+$, OH$^+$, and H$_2$O$^+$, are { hard to detect} and require { the presence of} background massive stars { for} absorption measurements. In this work, we propose { an alternative method to infer the CRIR in diffuse clouds using} the abundance ratios of OH/CO and HCO$^+$/CO. We have analyzed the response of chemical abundances of CO, OH, and HCO$^+$ on various environmental parameters of { the interstellar medium in diffuse clouds} and { found that their abundances are proportional to $\zeta_2$}. Our { analytic expressions} give an excellent calculation of the abundance of OH for $\zeta_2$ $\leq$10$^{-15}$\,s$^{-1}$, which { are potentially useful for modelling chemistry in hydrodynamical simulations}. The abundances of OH and HCO$^+$ were found to monotonically decrease with increasing density, while the CO abundance shows the opposite trend. 
{ With high-sensitivity absorption transitions of both CO (1--0) and (2--1) lines from ALMA, we have derived the H$_2$ number densities ($n_{\rm H_2}$) toward 4 line-of-sights (LOSs); assuming a kinetic temperature of $T_{\rm k}=50\,{\rm K}$, we find a range of (0.14$\pm$0.03--1.2$\pm$0.1)$\times$10$^2$\,cm$^{-3}$}. 
By comparing the observed and modelled HCO$^+$/CO ratios, we find that $\zeta_2$ in our diffuse gas sample is in the { range of $1.0_{-1.0}^{+14.8}$ $\times$10$^{-16}- 2.5_{-2.4}^{+1.4}$ $\times$10$^{-15}$\,s$^{-1}$}. { This is $\sim$2} times higher than the { average} value measured at higher extinction, supporting an attenuation of CRs as suggested by theoretical models.

\end{abstract}

\keywords{Interstellar medium(847) --- Interstellar molecules(849) --- Chemical abundances(224) --- Molecular clouds(1072)}


\section{Introduction} \label{sec:intro}

In the environments of { the} cold neutral { component of the interstellar medium (ISM),} where stellar photons cannot penetrate, { the low-energy (0.1 $<$ $E$ $<$ 1\,GeV)} cosmic-rays (CRs) { play a} critical { role in determining the ionization degree, in controlling the thermal balance and for initiating the chemistry} at high optical depths \citep{Dalgarno2006,Padovani2013,Vaupre2014,Grenier2015}. 
The cosmic-ray ionization rate (CRIR; $\zeta_2$\footnote{Throughout the text, $\zeta_2$ denotes the CRIR of molecular hydrogen.}) has been measured during the last few decades \citep{Spitzer1968,Webber1998,Dalgarno2006} { with the measured value to strongly depend on the adopted methodology} \citep{Dalgarno2006,Indriolo2012,Indriolo2015,Bacalla2019}. 
Various observations found that the CRIR is higher toward the Galactic center \citep[e.g., CMZ, $\zeta_2$ = 10$^{-15}$ $\sim$ 10$^{-14}$\,s$^{-1}$,][]{Oka2005,Le2016} and supernova remnants \citep[e.g., IC443, W49B, $\zeta_2$ $\sim$ 2$\times$10$^{-15}$\,s$^{-1}$,][]{Indriolo2010,Zhou2022} than that of nearby molecular clouds \citep[a few 10$^{-18}$ $\sim$ 10$^{-16}$\,s$^{-1}$,][]{Caselli1998,Indriolo2012,Neufeld2017}. Due to the interaction between CR particles and the ISM, the low-energy CRs attenuate { while propagating at higher column densities} \citep{Strong2007,Padovani2018,Padovani2020}. In low-density diffuse clouds (e.g., $n_{\rm H}$ $\sim$ a few hundred cm$^{-2}$), CRIR is -on average- almost an order of magnitude higher than that in dense clouds \citep{Indriolo2012}.

While $\zeta_2$ cannot be easily observed directly, { the use of tracers is favored instead}.
H$^+_3$ is one of the most commonly used tracers of the CRIR. It is produced by the CR ionization of H$_2$ and it is destroyed through reactions with abundant neutral species (e.g., CO, O) and electrons \citep{McCall1999,Geballe1999, Dalgarno2006,Indriolo2012}. The CRIR can be derived once the gas temperature, volume density ($n_{\rm H}$), and column densities (e.g. H$_2$, CO) are known. Other molecules that are directly relevant to the H$^+_3$ chemistry are considered as potential probes of the CRIR, such as HCO$^+$ and DCO$^+$ \citep{Guelin1982, van2000, Caselli1998}, OH$^+$ \citep{Indriolo2015,Bacalla2019}, and H$_2$O$^+$ \citep{Gerin2010b, Neufeld2017, Bialy2019}. { Measuring $\zeta_2$ with ions such as H$^+_3$, OH$^+$, and H$_2$O$^+$, requires the presence of bright massive stars in the background, which is not very common.} 
Furthermore, the deuterium species can only be detected in high extinction regions due to its relatively low abundance. { Although the above methodology can provide reasonable measurements, it is difficult and somehow impossible for general use.} 

{ However,} oxygen-bearing molecules have -in principle- the potential to constrain the CRIR; most of the formation of oxygen-bearing species (e.g., OH, HCO$^+$) starts from the hydrogenation of ionized oxygen (O$^+$). Since the ionization potential of atomic oxygen (13.62 eV) is { very close to} that of atomic hydrogen (13.6 eV), the majority of oxygen in the cold neutral medium (CNM) is in its atomic form. Thus, { in such FUV-shielded regions,} O$^+$ and oxygen-bearing species are the indirect products of CRs.

OH has long been proposed to be an alternative tracer of molecular gas due to its fairly constant abundance in the ISM \citep{Liszt1998,Liszt2002, Xu2016b, Li2018}. { The thermal emission and absorption lines of} OH 18\,cm have been detected extensively toward the CNM, which is extended to the outskirts of the molecular clouds and where the CO emission is faint or undetectable \citep{Turner1979, Magnani1988, Wannier1993, Cotten2012, Li2018, Busch2021}. HCO$^+$, in the traditional sense, is a dense gas tracer in the ISM. However, recent observations found that HCO$^+$ is ubiquitous in diffuse and translucent clouds \citep{Pety2017,Luo2020}, { especially with absorption measurements against strong continuum sources (e.g., quasars, H\,{\sc ii} regions)}. 

{ In this paper, we combine absorption observations of HCO$^+$, and absorption and emission observations of CO to calculate their column densities along each LOS.} We attempt to find the chemical connection between these oxygen-bearing molecules (CO, OH, and HCO$^+$) that are ubiquitous in diffuse clouds. We investigate the variance of chemical abundances of oxygen-bearing molecules under different environmental parameters (e.g., gas volume density, FUV intensity, CRIR), especially { the potential connection of their molecular abundance ratios as a probe of CRIR}. 

This paper is organized as follows. The observations and archival data used in this work are presented in Section \ref{sec:obs}. We present the results of column densities and constraints on the gas density in Section \ref{sec:results}. In Section \ref{sec:models}, we perform photodissociation region (PDR) modelling of CO, OH, and HCO$^+$ under various environmental parameters. We discuss the chemistry of CO, OH, and HCO$^+$ in the diffuse cloud and derive the chemical abundances in chemical equilibrium in Section \ref{sec:discussion}. We derive the abundance ratios of OH/CO and HCO$^+$/CO and constrain the CRIR by combining the observed HCO$^+$-to-CO abundance ratio and chemical models in the low-density ($n_\mathrm{H}$ $\sim$ 10$^2$\,cm$^{-3}$) diffuse cloud in Section \ref{sec:abundance ratio}. The main results and conclusions are summarized in Section \ref{sec:conclusion}.

\section{Observations and Data}\label{sec:obs}

{ \subsection{HCO$^+$ (1--0) and CO (1--0)}\label{sec:co10}}

The HCO$^+$ (1--0) and CO (1--0) absorption observations toward 13 strong continuum sources were carried out during Apr. 2016 to May. 2016 with ALMA (project ID: 2015.1.00503.S, PI: L. Bronfman). The calibration of the raw data was performed using the Common Astronomy Software Applications \citep{McMullin2007}. Self-calibration was performed toward 2 sources (3C454.3 and 3C120) to increase the signal-to-noise ratio and eliminate the spectral contamination from bandpass calibrators. For each line-of-sight (LOS), we decompose the absorption spectra ($\tau_\nu$) into different Gaussian components to derive the column density at each velocity component. 
A detailed description of observations and data reduction can be found in \citet{Luo2020}.

{ The HCO$^+$ (1--0) integrated optical depth (in units of km\,s$^{-1}$) and CO (1--0) emission toward another 15 sources are taken from Table 1 in \citet{Liszt2023}. The original data of  HCO$^+$ absorption was presented in \citet{Lucas1996}, \citet{Liszt2000}, and \citet{Liszt2018}. We exclude sources in which the brightness temperature of CO is bright ($T_{\rm mb}$ $\geq$ 5\,K) since we only focus on diffuse LOSs. }

{ 
\subsection{CO (2--1) data}\label{sec:co21}

The CO (2--1) absorption spectra toward 4 sources (3C454.3, 0607-157, 1730-130, and 1741-038) are taken from a blind survey of the absorption lines of bandpass calibrators in ALMA regular observations (Luo et al. 2023, in prep). The data reduction follows the same procedures as that of CO (1--0).

\subsection{Reddening $E$(B-V) data}\label{sec:ebv}

The $E$(B-V) data is taken from \citet{Green2019}, in which the values were derived by combining stellar photometry from Pan-STARRS 1 and 2MASS, and parallaxes from {\it Gaia}. We use the $E$(B-V) values to derive the total gas column density along the LOS.}

{ 
\section{Results}\label{sec:results}

\subsection{Constraints on the gas densities}\label{sec:nh}

For those components where both CO (1--0) and (2--1) transitions are available, we use the non-LTE radiative transfer code {\sc radex} \citep{Van2007} to constrain the gas densities along the LOSs. 
We account for H$_2$ as the main colliding partner of CO. 
We perform several models covering H$_2$ volume densities of 10$^1$ $\leq$ $n_{\rm H_2}$ $\leq$ 10$^4$\,cm$^{-3}$, CO column densities of 10$^{11}$ $\leq$ $N_{\rm CO}$ $\leq$ 10$^{16}$\,cm$^{-2}$ and kinetic temperatures of $10^1\leq T_{\rm k}\leq10^2\,{\rm K}$. 
We then find the optimum solutions by maximizing the likelihood function using the Markov chain Monte Carlo (MCMC) method, which is encoded in $emcee$ \citep{Foreman2013}. The likelihood function is defined as:
\begin{equation}
\rm ln \ p = -\frac{1}{2} \sum_i \left [ \frac{\left ( \tau_{obs}^i - \tau_{model}^i \right )^2}{{\sigma^i_{obs}}^2} + ln \left ( 2\pi {\sigma^i_{obs}}^2 \right ) \right ],
\end{equation}
where the ${\rm \tau_{obs}^i}$ and ${\rm \sigma^i_{obs}}$ are the optical depth and uncertainty of the observed {\it i}-th transition, respectively. ${\rm \tau_{model}^i}$ is the modelled optical depth by {\sc radex}.

The representative optimum results of $n_{\rm H_2}$ under different $T_{\rm k}$ are shown in the left panel of Fig.~\ref{fig:tk_nh}. As $T_{\rm k}$ increases by a factor of 10, the resultant $n_{\rm H_2}$ decreases by a factor of 2$\sim$5. However, the column densities are less influenced by $T_{\rm k}$ and vary by $<$4\% (for CO) and $<$0.1\% (for HCO$^+$) (with respect to the values at $T_{\rm k}$=50\,K, as shown in the middle and right panels of Fig.~\ref{fig:tk_nh}.) 

\begin{figure*}
\includegraphics[width=1.0\linewidth]{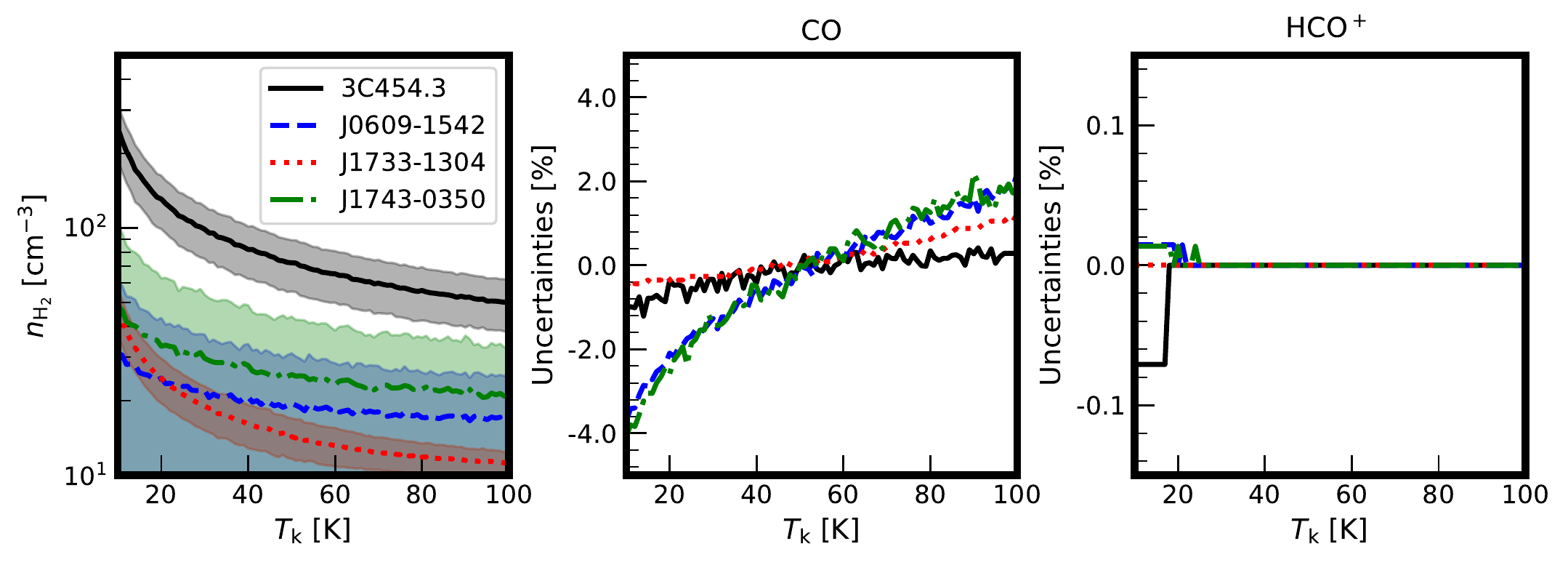}
\caption{Left: The optimum $n_{\rm H_2}$ solutions at specific $T_{\rm k}$. Different line curves denote results from different sources. Shaded areas represent 1$\sigma$ uncertainties. Middle: Dependence of relative uncertainties (with respect to the values at $T_{\rm k}$=50\,K) of column densities of CO on the $T_{\rm k}$. Right: The same as the middle panel but for HCO$^+$.}
\label{fig:tk_nh}
\end{figure*}

Table \ref{tab:nh} summarises the derived $n_{\rm H_2}$ and $n_{\rm H}$ of each velocity component at $T_{\rm k}$ = 10, 50, and 100\,K. The derived $n_{\rm H_2}$ is in the range of (0.11$\pm$0.01--4.4$\pm$0.4)$\times$10$^{2}$\,cm$^{-3}$. 

\begin{deluxetable*}{lccccccc}
\tablenum{1}
\tablecaption{Optical depths of CO (1--0) and (2--1); H$_2$ density derived from MCMC runs. \label{tab:nh}}
\tablewidth{700pt}
\tabletypesize{\scriptsize}
\tablehead{
\colhead{Sources}& \colhead{Velocity} & \multicolumn{2}{c}{$\tau_{\rm CO}$} & & \multicolumn{3}{c}{$n_{\rm H_2}$/10$^2$\,cm$^{-3}$} \\
\cline{3-4} \cline{6-8}
\colhead{} & \colhead{km\,s$^{-1}$} & \colhead{1--0} & \colhead{2--1} & & \colhead{$T_{\rm k}$=10} & \colhead{50} & \colhead{100\,K} 
}
\startdata
& -10.46$\pm$0.04 & 0.055$\pm$0.002 & 0.027$\pm$0.002 & & 2.5$\pm$0.7 & 0.7$\pm$0.2 & 0.5$\pm$0.1 \\
3C454.3 & -9.47$\pm$0.01 & 0.419$\pm$0.004 & 0.248$\pm$0.007 & & 3.6$\pm$0.3 & 1.0$\pm$0.1 & 0.68$\pm$0.04  \\
& -8.94$\pm$0.01 & 0.149$\pm$0.003 & 0.092$\pm$0.004 & & 4.4$\pm$0.4 & 1.2$\pm$0.1 & 0.81$\pm$0.07  \\
J0609-1542 & 7.37$\pm$0.02 & 0.12$\pm$0.02 & 0.032$\pm$0.007 & & 0.3$\pm$0.3 & 0.2$\pm$0.1 & 0.18$\pm$0.09  \\
J1733-1304 & 5.03$\pm$0.01 & 1.55$\pm$0.04 & 0.553$\pm$0.004 & & 0.46$\pm$0.10 & 0.14$\pm$0.03 & 0.11$\pm$0.01  \\
J1743-0350 & 3.89$\pm$0.05 & 0.17$\pm$0.01 & 0.06$\pm$0.02 & & 0.49$\pm$0.53 & 0.26$\pm$0.18 & 0.21$\pm$0.12  \\
& 5.6$\pm$0.5 & 0.065$\pm$0.007 & 0.039$\pm$0.008 & & 3.2$\pm$2.6 & 0.9$\pm$0.6 & 0.6$\pm$0.4 \\
\enddata
\end{deluxetable*}

\subsection{Calculation of column densities}\label{sec:col}

For the rest of the components where only CO (1--0) and HCO$^+$ (1--0) are available, we consider an excitation temperature of $T_{\rm ex}=4\,{\rm K}$ for CO and $T_{\rm ex}=2.73\,{\rm K}$ for HCO$^+$ \citep[e.g.][]{Liszt1996,Godard2010,Luo2020}.

The column densities are calculated with \citep{Mangum2015}:
\begin{equation}
N_{tot} = \frac{3h}{8{\pi}^3\left | \mu_{lu} \right |^2} \frac{Q_{rot}}{g_{u}} \frac{e^{\frac{E_u}{kT_{ex}}}} {e^{\frac{h\nu}{kT_{ex}}}-1}  \int \tau_\nu d\upsilon,
\label{eq:n_tot}
\end{equation}
where $\mathrm{\left | \mu_{lu} \right |^2}$ is the dipole matrix element, $\mathrm{Q_{rot}}$ is the rotational partition function, $\mathrm{g_u}$ is the degeneracy of the upper energy level, and $\mathrm{E_u}$ is the energy of the upper energy level.
For each transition, the $\mathrm{\left | \mu_{lu} \right |^2}$, $\mathrm{E_u}$, and the rest frequency $\nu$ are taken from the CDMS database \citep{Muller2001,Muller2005}. For linear molecules, the partition function is given by \citep{McDowell1987}:
\begin{equation}
Q_{tot} = \frac{kT_{\rm ex}}{hB_0} e^{\frac{hB_0}{3kT_{\rm ex}}} ,
\label{eq:q}
\end{equation}
where $B_0$ is the rigid rotor rotation constant.
The calculated column densities of CO and HCO$^+$ are summarised in Table \ref{tab:col}.

\begin{deluxetable*}{lccccc}
\tablenum{2}
\tablecaption{The $E$(B-V) values, column densities of CO and HCO$^+$, fit parameters from {\sc 3d-pdr} models (see Section \ref{sec:CRIR}). \label{tab:col}}
\tablewidth{700pt}
\tabletypesize{\scriptsize}
\tablehead{
\colhead{Sources}& \colhead{$E$(B-V)} & \colhead{$N_{\rm CO}$} & \colhead{$N_{\rm HCO^+}$} & \multicolumn{2}{c}{Fit parameters} \\
\cline{2-6}
\colhead{} & \colhead{mag}  & \colhead{10$^{14}$\,cm$^{-2}$} & \colhead{10$^{11}$\,cm$^{-2}$} & \colhead{$n_{\rm H}$/10$^2$\,cm$^{-3}$} & \colhead{$\zeta_2$/10$^{-15}$\,s$^{-1}$}
}
\startdata
3C454.3 V1 & 0.106$\pm$0.002 & 0.84$\pm$0.03 & 1.41$\pm$0.03 & $1.0_{-0.7}^{1.0}$ & $1.85_{-1.66}^{0.30}$(*) \\ 
3C454.3 V2 & 0.106$\pm$0.002 & 3.35$\pm$0.03 & 3.54$\pm$0.09 & $1.0_{0.0}^{0.0}$ & $1.58_{-0.26}^{0.26}$(*) \\ 
3C454.3 V3 & 0.106$\pm$0.002 & 1.03$\pm$0.02 & 1.09$\pm$0.14 & $1.3_{-0.6}^{1.9}$ & $0.12_{-0.02}^{2.81}$(*) \\ 
J0609-1542 & 0.203$\pm$0.004 & 0.52$\pm$0.07 & 6.82$\pm$0.21 & $0.1_{0.0}^{0.0}$ & $0.14_{-0.02}^{0.05}$ \\ 
J1733-1304 & 0.513$\pm$0.009 & 11.40$\pm$0.33 & 17.41$\pm$0.49 & $0.3_{0.0}^{0.0}$ & $0.34_{-0.06}^{0.06}$ \\ 
J1743-0350 V1 & 0.530$\pm$0.008 & 2.79$\pm$0.19 & 7.34$\pm$1.23 & $0.3_{-0.2}^{0.4}$ & $0.25_{-0.18}^{0.38}$(*) \\ 
J1743-0350 V2 & 0.530$\pm$0.008 & 2.28$\pm$0.24 & 4.97$\pm$0.56 & $0.5_{-0.4}^{0.3}$ & $0.25_{-0.20}^{0.75}$(*) \\ 
3C120 & 0.265$\pm$0.006 & 1.14$\pm$0.24 & 2.71$\pm$0.58 & $0.8_{-0.7}^{0.8}$ & $1.17_{-1.14}^{0.42}$(*) \\ 
J1745-0753 V1 & 0.672$\pm$0.008 & 2.18$\pm$0.89 & 1.27$\pm$0.87 & $3.2_{-3.1}^{28.5}$ & $0.86_{-0.86}^{9.14}$(*) \\ 
J1745-0753 V2 & 0.672$\pm$0.008 & 8.06$\pm$1.26 & 5.60$\pm$2.46 & $1.0_{-0.9}^{2.2}$ & $0.86_{-0.81}^{3.12}$(*) \\ 
J1745-0753 V3 & 0.672$\pm$0.008 & 0.11$\pm$0.32 & 2.22$\pm$1.20 & $0.2_{-0.1}^{1.8}$ & $0.10_{-0.10}^{1.48}$(*) \\ 
J0211+1051 & 0.097$\pm$0.006 & 5.71$\pm$0.40 & 8.43$\pm$0.32 & $0.5_{0.0}^{0.0}$ & $0.54_{-0.09}^{0.09}$ \\ 
J0325+2224 & 0.239$\pm$0.008 & 14.92$\pm$0.81 & 11.21$\pm$0.19 & $0.3_{0.0}^{0.0}$ & $0.34_{-0.06}^{0.06}$ \\ 
J0356+2903 & 0.159$\pm$0.003 & 25.71$\pm$0.67 & 16.64$\pm$1.00 & $31.6_{-15.8}^{0.0}$ & $0.34_{-0.34}^{0.06}$ \\ 
J0401+0413 & 0.230$\pm$0.009 & 3.02$\pm$0.76 & 5.44$\pm$0.23 & $0.2_{-0.1}^{0.6}$ & $0.34_{-0.26}^{0.52}$(*) \\ 
J0403+2600 & 0.141$\pm$0.002 & 9.84$\pm$1.06 & 6.66$\pm$0.32 & $0.6_{0.0}^{0.2}$ & $0.74_{-0.10}^{0.12}$(*) \\ 
J0406+0637 & 0.248$\pm$0.006 & 10.00$\pm$0.63 & 5.99$\pm$0.57 & $0.8_{-0.2}^{0.2}$ & $1.00_{-0.37}^{0.17}$(*) \\ 
J0407+0742 & 0.159$\pm$0.006 & 3.02$\pm$0.67 & 5.88$\pm$0.34 & $0.4_{-0.2}^{0.2}$ & $0.63_{-0.47}^{0.10}$(*) \\ 
J0426+2327 & 0.292$\pm$0.008 & 76.82$\pm$0.71 & 28.51$\pm$0.63 & $31.6_{-15.8}^{0.0}$ & $0.18_{-0.10}^{0.11}$ \\ 
J0427+0457 & 0.292$\pm$0.017 & 6.19$\pm$0.71 & 6.88$\pm$0.27 & $0.3_{-0.2}^{0.3}$ & $0.54_{-0.38}^{0.19}$(*) \\ 
J0437+2037 & 0.451$\pm$0.006 & 10.63$\pm$0.41 & 17.09$\pm$0.81 & $0.3_{0.0}^{0.0}$ & $0.34_{-0.06}^{0.06}$ \\ 
J0431+1731 & 0.354$\pm$0.004 & 11.11$\pm$0.78 & 11.21$\pm$1.22 & $0.5_{-0.1}^{0.0}$ & $0.54_{-0.14}^{0.09}$(*) \\ 
J0440+1437 & 0.442$\pm$0.005 & 13.17$\pm$0.46 & 13.42$\pm$0.34 & $0.4_{-0.1}^{0.0}$ & $0.46_{-0.12}^{0.07}$ \\ 
J0449+1121 & 0.398$\pm$0.009 & 3.65$\pm$0.52 & 7.21$\pm$0.24 & $0.2_{-0.1}^{0.2}$ & $0.29_{-0.19}^{0.17}$(*) \\ 
J0502+1338 & 0.371$\pm$0.003 & 39.04$\pm$0.49 & 20.08$\pm$0.65 & $31.6_{0.0}^{0.0}$ & $10.00_{-1.61}^{1.61}$ \\ 
J0510+1800 & 0.212$\pm$0.008 & 2.70$\pm$0.57 & 1.44$\pm$0.06 & $0.8_{-0.5}^{2.4}$ & $2.51_{-2.35}^{1.47}$(*) \\  
\enddata
\tablenotetext{a}{V1, V2, and V3 represent different velocity components along the LOSs.}
\tablenotetext{}{Symbol ``*'' represents high confidence fit from {\sc 3d-pdr} models (see Section \ref{sec:CRIR}).}
\end{deluxetable*}
}

\section{Photodissociation region modelling}\label{sec:models}

{ Variations in} chemical abundances { can} be used as a diagnostic tool for { estimating the environmental parameters of the ISM}. 
To better understand the response of the abundances of { the molecular species we examine} (CO, OH, HCO$^+$) and which are ubiquitously detected in diffuse and translucent clouds, we perform chemical simulations under a range of { different} environmental parameters (e.g. varying densities, FUV intensities, and cosmic-ray ionization rates). For these calculations, we use the publicly available code {\sc 3d-pdr}\footnote{https://uclchem.github.io/3dpdr/} \citep{Bisbas2012}.

In our modelling, we use a suite of one-dimensional uniform slabs with densities of $n_{\rm H}=10^m\,{\rm cm}^{-3}$ where $m=2.0,2.5,3.0,3.5$, interacting with four FUV intensities $\chi$/$\chi_0$=1,10,30,100 (where $\chi$ is normalized to the { spectrum} of \citealt{Draine1978}), and four CRIR $\zeta_2=(5,10,50,100)\times$10$^{-17}\,{\rm s}^{-1}$. The FUV radiation field is considered as plane-parallel that impinges from one direction. The diffuse component of radiation is not accounted for. The maximum value of $A_\mathrm{V}$ in our simulations is 20\,mag. { The gas-phase element abundances relative to hydrogen are shown in Table \ref{tab:abundance}. We set all of the element carbon in the ionized phase with an abundance of 1.4$\times$10$^{-4}$ and 60\% of the total hydrogen in the molecular phase as the initial conditions in our PDR models \citep[see][for further details]{Roellig07}.} Throughout the text, the diffuse cloud is referred to as molecular gas with $n_{\rm H}$ $\leq$500\,cm$^{-3}$ and $A_{\rm V}$ $\leq$1\,mag, the translucent cloud is referred to 500$\leq$ $n_{\rm H}$ $\leq$5000\,cm$^{-3}$ and $A_{\rm V}$ $\leq$5\,mag.

\begin{deluxetable}{cccc}
\tablenum{3}
\tablecaption{The gas phase elemental abundances relative to hydrogen in the PDR models. \label{tab:abundance}}
\tablewidth{700pt}
\tabletypesize{\scriptsize}
\tablehead{
\colhead{Elements}& \colhead{Abundance} & \colhead{Elements} & \colhead{Abundance}
}
\startdata
C$^+$  &  1.4$\times$10$^{-4}$ & H$_2$  &  3$\times$10$^{-1}$   \\
He  &  1$\times$10$^{-1}$ & H  &  4$\times$10$^{-1}$ \\
O  &  3$\times$10$^{-4}$  & \\
\enddata
\end{deluxetable}

\begin{figure*}
\includegraphics[width=1.0\linewidth]{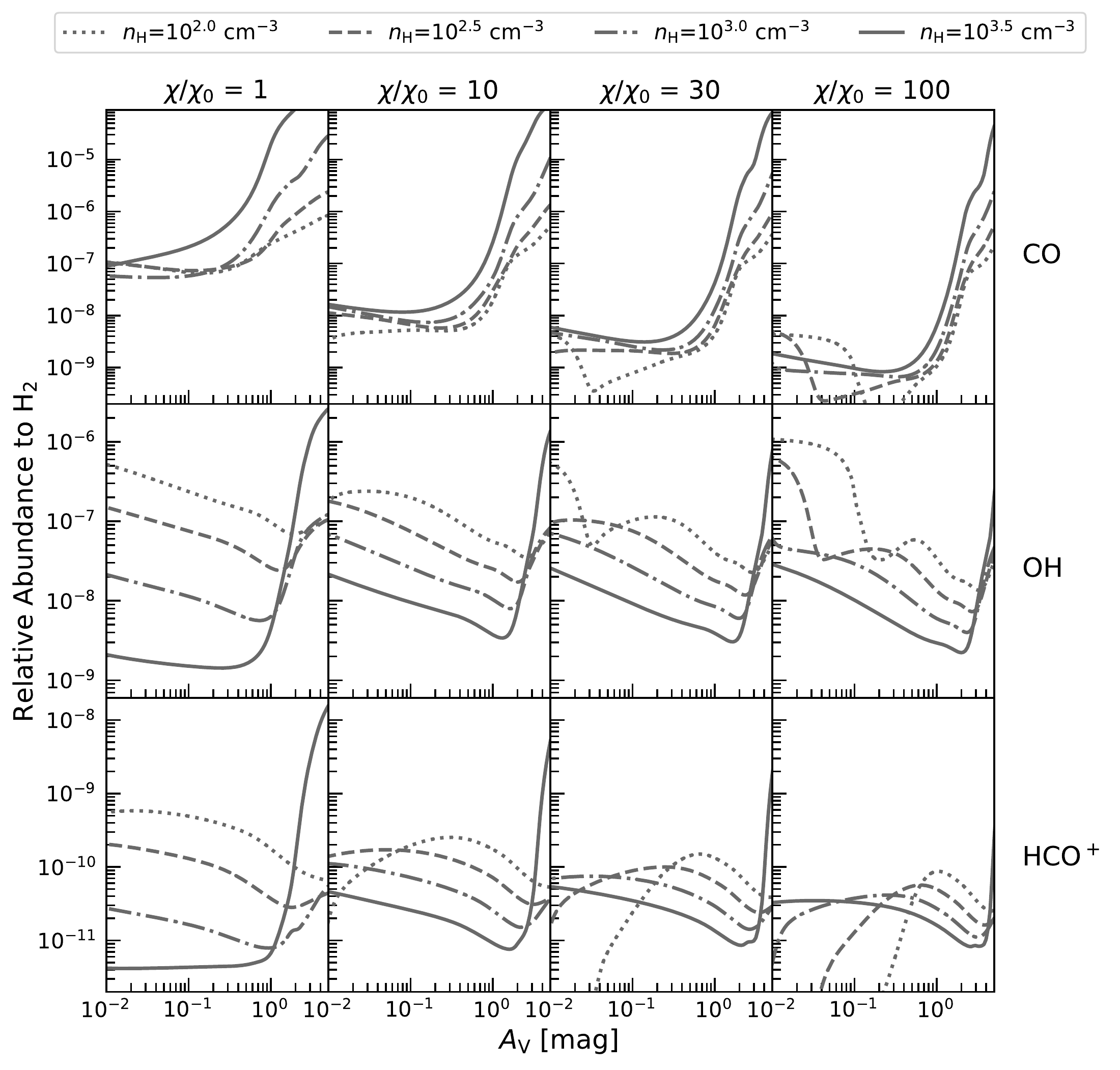}
\caption{The chemical abundances of CO, OH, and HCO$^+$ { as a function of $A_{\rm V}$} using {\sc 3d-pdr} for four densities ($n_\mathrm{H}$ = 10$^2$, 10$^{2.5}$, 10$^3$, and 10$^{3.5}$\,cm$^{-3}$) and four FUV intensities ($\chi$/$\chi_0$ = 1, 10, 30, 100) for a fixed CRIR of $\zeta_2=10^{-16}\,{\rm s}^{-1}$.}
\label{fig:pdr_den}
\end{figure*}

Figure \ref{fig:pdr_den} shows our simulation results for a fixed $\zeta_2=10^{-16}\,{\rm s}^{-1}$, { in which the relative abundance of each species to H$_2$ is defined as the ratio of 
the corresponding column densities along the LOS.}
The variance of CO abundance is shown in the first row of Fig.~\ref{fig:pdr_den}. CO is mostly photodissociated, especially at lower extinction regions (e.g., $A_{\rm V}$ $<$1\,mag). As a consequence, the abundance of CO at low $A_{\rm V}$ decreases by two orders of magnitude (from $\sim$ 10$^{-7}$ to $\sim$ 10$^{-9}$) as the FUV intensity increases from $\chi$/$\chi_0$=1 to 100. The abundance of CO increases with increasing $A_{\rm V}$ (e.g., 0.6$\leq$ $A_{\rm V}$ $\leq$5\,mag) due to FUV shielding by dust and also self-shielding. However, the abundance of CO slightly decreases by a factor of a few from low to intermediate extinctions ($A_{\rm V}$ $\leq$0.6\,mag), which is due to the decrease of the abundance of its precursors (see Section \ref{sec:discussion}). For a fixed FUV intensity, the abundance of CO increases with increasing density. The divergence of CO abundance at different densities is smaller (within a factor of a few, from 5$\times$10$^{-8}$ to 5$\times$10$^{-7}$ for $\chi$/$\chi_0$ = 1) at low extinction ($A_{\rm V}$ $<$1\,mag) but larger, over an order of magnitude, at high $A_{\rm V}$.  At high-density or low FUV intensity models, the abundance of CO reaches the canonical value (e.g., $\sim$10$^{-4}$ as in the solar neighborhood) at lower extinction ($A_{\rm V}$ = 2\,mag for $\chi$/$\chi_0$ = 1 and $n_{\rm H}$ = 10$^{3.5}$\,cm$^{-3}$) than that of low-density or high FUV intensity models ($A_{\rm V}$ = 5\,mag for $\chi$/$\chi_0$ = 10 and $n_{\rm H}$ = 10$^{3.5}$\,cm$^{-3}$). 

On the contrary, the abundance of OH has a different { pattern} to that of CO, as shown in the second row of Fig.~\ref{fig:pdr_den}. At low $A_{\rm V}$ (e.g., $\lesssim$2\,mag), the abundance of OH increases with the increasing FUV intensity, especially for high-density simulations (e.g., $n_{\rm H}$ = 10$^{3.5}$\,cm$^{-3}$). { It increases} from $\sim$10$^{-9}$ to 10$^{-8}$ as the FUV intensity increases from $\chi$/$\chi_0$=1 to 100. The abundance of OH decreases with the increasing $A_{\rm V}$ when $A_{\rm V}$ $<$2\,mag. The slope of the decreasing trend of OH abundance with $A_{\rm V}$ becomes larger for lower densities and higher FUV intensities. As can be seen from Fig.~\ref{fig:pdr_den}, the abundance of OH decreases by a factor of 2 for { the} high-density ($n_{\rm H}$ = 10$^{3.5}$\,cm$^{-3}$) and low FUV intensity ($\chi$/$\chi_0$=1) model, while it decreases by two orders of magnitude for { the} low-density ($n_{\rm H}$ = 10$^{2}$\,cm$^{-3}$) and high FUV intensity ($\chi$/$\chi_0$=100) model. 

The knee point where the OH abundance trend turns, is approximately at an $A_{\rm V}$ of 2\,mag, after which { it} monotonically decreases with the increase of density below the $A_{\rm V}$ value, while it converges above the $A_{\rm V}$ value (except for $n_{\rm H}$ = 10$^{3.5}$\,cm$^{-3}$). Particularly, the divergence of OH abundance at low FUV intensities is larger than that of at high FUV intensity. With the density increase from 10$^2$\,cm$^{-3}$ to 10$^{3}$\,cm$^{-3}$, the abundance of OH decreases by nearly two orders of magnitude at $\chi$/$\chi_0$=1, while it decreases only less than one order of magnitude at $\chi$/$\chi_0$=100. 

{ Furthermore,} the abundance of HCO$^+$ follows a similar { trend to that of} OH at low FUV intensity; HCO$^+$ decreases with { the increase of} density and $A_{\rm V}$ ($A_{\rm V}$ $<$2\,mag). While { this} trend is maintained for the high-density (e.g., $n_{\rm H}$ $>$10$^3$\,cm$^{-3}$) and low FUV intensity models (e.g., $\chi$/$\chi_0$ $<$100), the abundance of HCO$^+$ declines at low extinction for low-density and high FUV intensity models. { At high FUV intensities, HCO$^+$ does not form as efficiently as OH does.}
The difference between HCO$^+$ and OH is most likely due to the interruption of the formation of HCO$^+$ through the OH channel at high FUV intensities (see Section \ref{sec:discussion}).

\begin{figure*}
\includegraphics[width=1.0\linewidth]{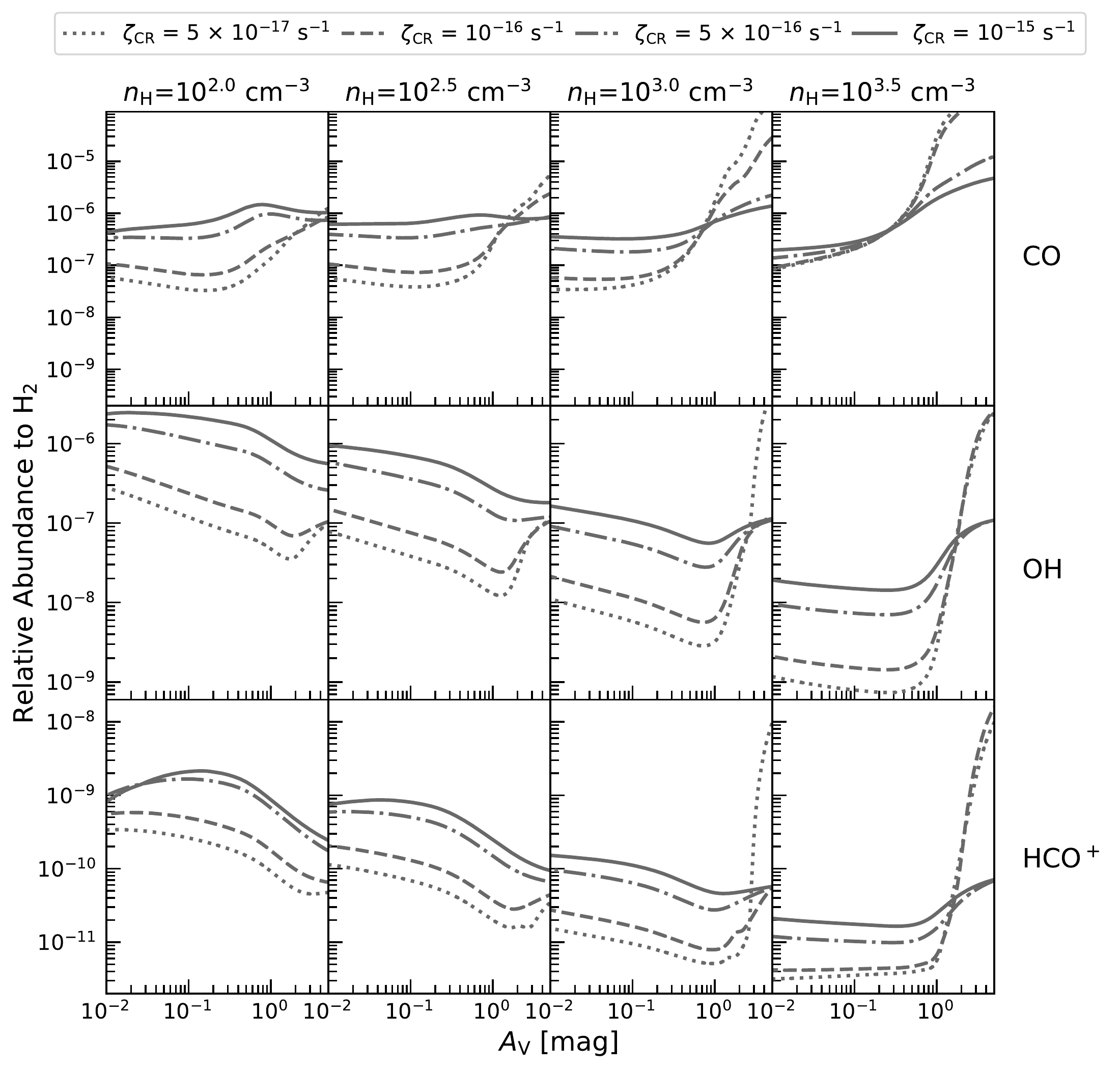}
\caption{As in Fig.~\ref{fig:pdr_den} but for a fixed FUV intensity of $\chi/\chi_0=1$.}
\label{fig:pdr_cr}
\end{figure*}

Figure \ref{fig:pdr_cr} shows our simulations with a constant FUV intensity ($\chi$/$\chi_0$=1). 
The abundances of CO, OH, and HCO$^+$ have a similar trend as $\zeta_2$ varies: they increase with increasing CRIR at low extinction (e.g., $A_{\rm V}$ $<$ 1\,mag for CO at $n_{\rm H}$=10$^3$\,cm$^{-3}$), while they have an inverse relation at high extinctions. 
The abundance of CO increases by an order of magnitude as the CRIR increase from 10$^{-16}$ to 10$^{-15}$\,s$^{-1}$ at low extinctions for low-density models (e.g., $n_{\rm H}$ $\leq$10$^3$\,cm$^{-3}$). The $A_{\rm V}$ value of the ``inverse point'' where the dependence of CO abundance on the CRIR turns opposite, is higher ($A_{\rm V}$ $\sim$5\,mag) for low-density models and lower ($A_{\rm V}$ $\sim$0.3\,mag) for high-density models. For the $n_{\rm H}$ = 10$^{3.5}$\,cm$^{-3}$ models, { an order of magnitude increase in CRIR (from $\zeta_2=10^{-16}\,{\rm s}^{-1}$ to $10^{-15}\,{\rm s}^{-1}$) affects the CO abundance from a factor of $\sim2$ at low $A_{\rm V}$ to $\gtrsim10^2$ at higher $A_{\rm V}$.}

{ As an important ISM environmental parameter,} the CRIR has a high impact on the abundance of OH. { In particular, it} is almost proportional to $\zeta_2$ at low $A_{\rm V}$. As seen in the second row of Fig.~\ref{fig:pdr_cr}, the abundance of OH increases by an order of magnitude as $\zeta_2$ increases from 10$^{-16}$ to 10$^{-15}$\,s$^{-1}$ at low $A_{\rm V}$ (e.g., $A_{\rm V}$ $<$1\,mag) for all four densities { explored}. The $A_{\rm V}$ value of the ``inverse point" is larger than that of CO, which decreases with increasing density. At high extinction ($A_{\rm V}$ $>$2\,mag) and high-densities ($n_{\rm H}$ $>$10$^3$\,cm$^{-3}$), the abundance of OH decreases with the increasing $\zeta_2$.

The dependence of HCO$^+$ abundance on the CRIR is similar to that of OH. With the CRIR increase by an order of magnitude, the abundance of HCO$^+$ increases by an order of magnitude at low $A_{\rm V}$ for $n_{\rm H}$ $\leq$10$^3$\,cm$^{-3}$. However, at low-density models ($n_{\rm H}$ = 10$^2$\,cm$^{-3}$), an increase of CRIR from 5$\times$10$^{-16}$\,s$^{-1}$ to 10$^{-15}$\,s$^{-1}$ only results in a slight increase ($\leq$30\%) on the abundance of HCO$^+$, meaning that the abundance of HCO$^+$ would hit a maximum value in this case (see Section \ref{sec:hcop abundance} for more discussion). At high-densities ($n_{\rm H}\sim$10$^{3.5}$\,cm$^{-3}$), the increase of CRIR by an order of magnitude increases HCO$^+$ abundance by a factor of $\sim$5. The $A_{\rm V}$ value of the ``inverse point" is the same as that of OH.

\section{Analysis and Discussion}\label{sec:discussion}

\subsection{The abundance of OH in diffuse cloud}\label{sec:oh abundance}

In diffuse clouds, the formation of OH can be traced back from two channels that form OH$^+$. The first one starts from the charge transfer reaction:
\begin{equation}
\rm H^+ + O \rightarrow O^+ + H \tag{R1},
\label{r:R1}
\end{equation}
%
%
{ with O$^+$ forming} OH$^+$ through reaction:
\begin{equation}
\rm O^+ + H_2 \rightarrow OH^+ + H \tag{R2}.
\label{r:R2}
\end{equation}

In addition to the reactions \ref{r:R1} and \ref{r:R2}, atomic oxygen can directly react with H$_3^+$ to form OH$^+$ (the second channel):
\begin{equation}
\rm O + H_3^+ \rightarrow OH^+ + H_2 \tag{R3}.
\label{r:R3}
\end{equation}

Once OH$^+$ has formed, it can hydrogenate to form the precursors of OH, namely H$_2$O$^+$ and H$_3$O$^+$:
\begin{align}
\rm OH^+ + H_2  \rightarrow &\ \rm H_2O^+ + H  \tag{R4}, \label{r:R4}\\ 
\rm  H_2O^+ + H_2 \rightarrow &\  \rm H_3O^+ + H \tag{R5},
\label{r:R5}
\end{align}
in which reaction \ref{r:R4} is the main destruction process for OH$^+$.
OH is formed through electron recombination reactions:
\begin{align}
\rm H_2O^+ + e  \rightarrow &\ \rm OH + H  \tag{R6}, \label{r:R6}\\ 
\rm   \rightarrow &\  \rm O + H_2 + H \tag{R7}, \label{r:R7}\\
\rm   \rightarrow &\  \rm O + H + H \tag{R8},
\label{r:R8}
\end{align}
and
\begin{align}
\rm H_3O^+ + e  \rightarrow &\ \rm OH + H + H \tag{R9}, \label{r:R9}\\ 
\rm   \rightarrow &\  \rm OH + H_2  \tag{R10}, \label{r:R10}\\
\rm   \rightarrow &\  \rm H_2O + H \tag{R11}.
\label{r:R11}
\end{align}
{ From reaction \ref{r:R11}, H$_2$O also contributes to} the formation of OH by photodissociation:
\begin{equation}
\rm H_2O + h\nu \rightarrow OH + H \tag{R12}.
\label{r:R12}
\end{equation}
Reaction \ref{r:R12} is also the main destruction path of H$_2$O. { The minor} destruction path of H$_2$O is through C$^+$ in diffuse clouds:
\begin{align}
\rm H_2O + C^+ \rightarrow HCO^+ + H \tag{R13}.
\label{r:R13}
\end{align}

The destruction of OH in diffuse clouds includes photodissociation:
\begin{equation}
\rm OH + h\nu \rightarrow O + H \tag{R14},
\label{r:R14}
\end{equation}
reaction with C$^+$:
\begin{align}
\rm OH + C^+ \rightarrow CO^+ + H \tag{R15}, \label{r:R15}\\
\rm OH + C^+ \rightarrow CO + H^+ \tag{R16},
\label{r:R16}
\end{align}
and reaction with H$^+$:
\begin{equation}
\rm OH + H^+ \rightarrow OH^+ + H \tag{R17}.
\label{r:R17}
\end{equation}

The OH abundance ($x$(OH)=$n$(OH)/$n_{\rm H}$) in chemical equilibrium can be solved when the formation and destruction processes reach a balance (see detailed derivation in Appendix \ref{sec:xco}). In diffuse clouds, photodissociation dominates the destruction of OH, and reactions \ref{r:R15} and \ref{r:R16} play a minor role. Since the abundance of C$^+$ is an order of magnitude higher than H$^+$ { when the $\zeta_2$ is not high \citep[e.g., $\zeta_2$ $\lesssim$10$^{-15}$\,s$^{-1}$,][]{LePetit2016}}, we ignore here the contribution from \ref{r:R17} (the last term in equation \ref{eq:doh1}). 
Thus, the abundance of OH can be written\footnote{In the following, $\rm k_{Rx}$ represents the reaction rate of Reaction Rx, and $\rm k_{pd}(Rx)$ represents the photodissociation rate of Reaction Rx.} as:
\begin{align}
\rm {\it x}(OH) =&\ \rm \frac{{\it x}(O)\zeta_2\theta}{k_{pd}(R14)+{\it n}(C^+)(k_{R15}+k_{R16})} \notag \\
&\ \rm \times [1-\frac{{\it n}(H_2)k_{R5}\delta+{\it n}(e)(k_{R7}+k_{R8})}{{\it n}(H_2)k_{R5}+{\it n}(e)(k_{R6}+k_{R7}+k_{R8})}] \notag \\
=&\ \rm \frac{{\it x}(O)\zeta_2\theta\phi}{k_{pd}(R14)+{\it n}(C^+)(k_{R15}+k_{R16})},
\label{eq:xoh1}
\end{align}
where $\phi$ is the term { in the square brackets} of equation \ref{eq:xoh1} and $\theta$ is in the form of equation \ref{eq:theta}.
Since the term $\delta$ is small ($\sim$0.1) in { average} diffuse ISM conditions (e.g., T =30\,K, $A_{\rm V}$ = 1\,mag), $\phi$ $\approx$1. To a good approximation, equation \ref{eq:xoh1} can be simplified as:
\begin{align}
\rm {\it x}(OH) =&\ \rm  \frac{{\it x}(O)\zeta_2\theta}{k_{pd}(R14)+{\it n}(H){\it x}(C^+)(k_{R15}+k_{R16})}.
\label{eq:xoh2}
\end{align}

Since there is an anti-correlation between {\it x}(OH) and $n_\mathrm{H}$ { as density increases}, the abundance of OH decreases monotonically ({ similar to that shown in} Fig.~\ref{fig:pdr_den}). On the other hand, if the UV radiation field increase or the extinction decreases, the gas temperature will increase. Therefore, all reaction rates in equation \ref{eq:xoh2} will decrease and, { therefore, the} abundance of OH will { also} increase. 

\begin{figure}
\includegraphics[width=1.0\linewidth]{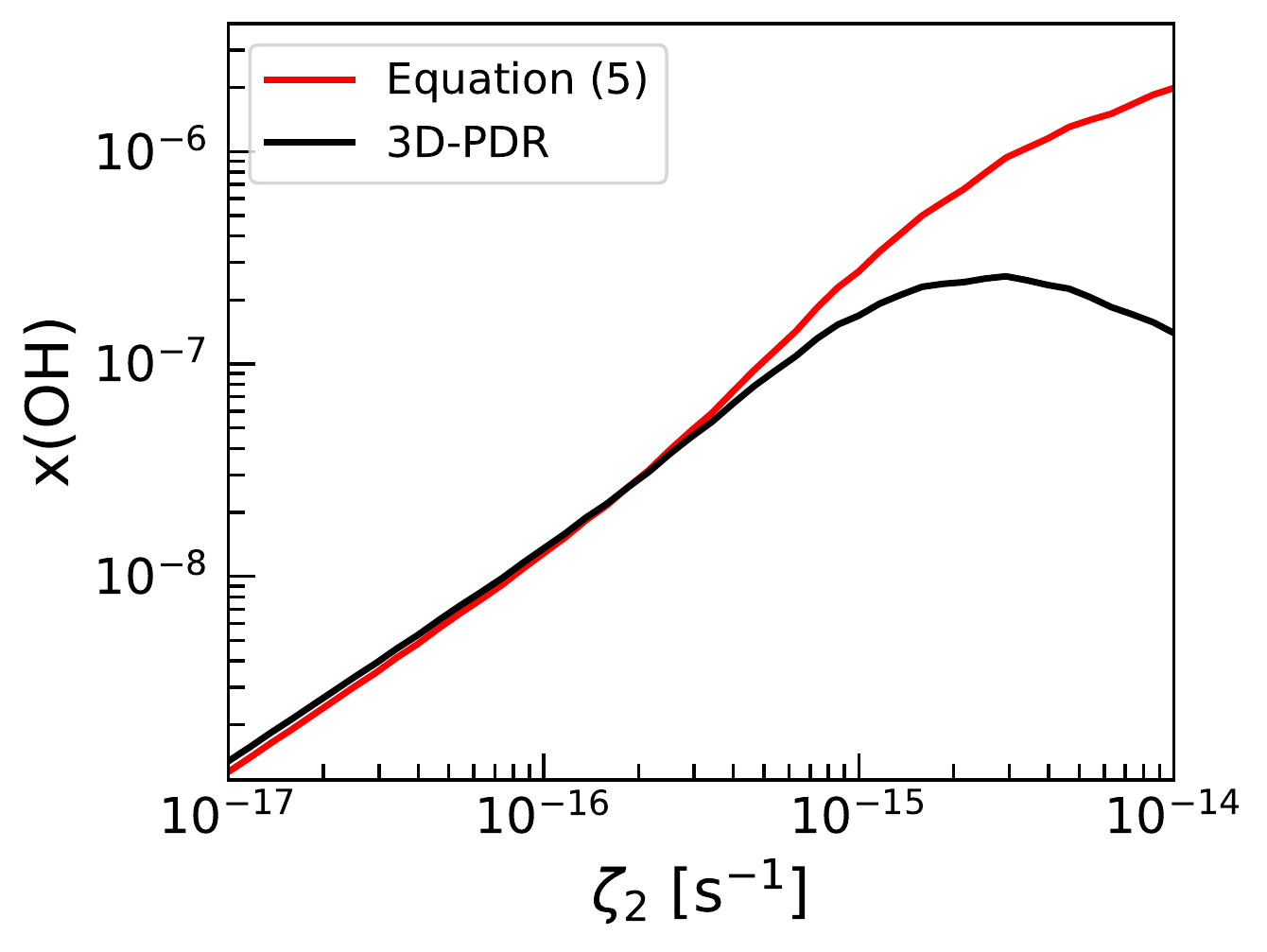}
\caption{The value of {\it x}(OH) as a function of CRIR, in which the red curve is obtained from equation \ref{eq:xoh2} and the black curve is from {\sc 3d-pdr} models. The extinction is at 1\,mag and the gas density is 10$^2$\,cm$^{-3}$.}
\label{fig:oh_cr}
\end{figure}

As { can be seen} from equation \ref{eq:xoh2}, the abundance of OH is proportional to $\zeta_2$ in the diffuse cloud. Figure~\ref{fig:oh_cr} { shows the} resultant abundance of OH ($A_{\rm V}$ = 1\,mag, $n_{\rm H}$ = 10$^2$\,cm$^{-2}$) from both equations \ref{eq:xoh2} { as well as the} {\sc 3d-pdr} models, in which the reaction rates are taken from \citet{McEl13}. { It should be emphasized} that equation \ref{eq:xoh2} can only be used { under average} ISM conditions. At high FUV radiation fields or very low $A_{\rm V}$, the main formation pathway of OH is contributed by neutral-neutral reactions through atomic O and H, which are not included in our analysis\footnote{The reaction rate of O and H$_2$ is $<<$10$^{-30}$\,cm$^{3}$\,s$^{-1}$ at a gas temperature below 100\,K, which can be ignored. However, once the temperature increases above 200\,K, the formation rate { increases to} 10$^{-20}$\,cm$^{3}$\,s$^{-1}$, which is comparable to that in equation \ref{eq:noh3}.}. 
Equation \ref{eq:xoh2} { represents} a good approximation ({ with} $<$60\% uncertainty) for $\zeta_2$ $<$10$^{-15}$\,s$^{-1}$ { when compared with {\sc 3d-pdr} models}. 

However, { in the PDR simulations,} the abundance of OH reaches the maximum at $\zeta_2$ $\sim$ 2$\times$10$^{-15}$\,s$^{-1}$. { For higher $\zeta_2$, the OH abundance decreases. This is} because the destruction of OH at high CRIR is dominated by H$^+$ (Reaction \ref{r:R17}) rather than by photodissociation and C$^+$ (Reactions \ref{r:R14}$\sim$\ref{r:R16}), which has been ignored in deriving equation \ref{eq:xoh2}. Similar work using isothermal simulations has been reported by \citet[][their Fig.~10]{Bisbas2017}, in which the OH abundance peaks at $\zeta_2$/$n_{\rm H}$ $\approx$ 2$\times$10$^{-17}$\,cm$^3$\,s$^{-1}$.

Though there are limitations, we still highlight the usage of equation \ref{eq:xoh2} at $\zeta_2$ $<$10$^{-15}$\,s$^{-1}$ instead of running chemical networks, especially when { coupled with} hydrodynamical simulations, { as this dramatically reduces the overall computational expense}.

\subsection{The abundance of HCO$^+$ in diffuse cloud}\label{sec:hcop abundance}

The formation of HCO$^+$ starts from  ion-neutral reactions:
\begin{align}
\rm CO^+ + H_2 \rightarrow &\ \rm HCO^+ + H \tag{R18},\label{r:R18}\\
\rm C^+ + H_2O \rightarrow &\ \rm HCO^+ + H \tag{R19},\label{r:R19}\\
\rm CH + O \rightarrow &\ \rm HCO^+ + e \tag{R20},
\label{r:R20}
\end{align}
where CO$^+$ is the result of reaction \ref{r:R15}. Reaction \ref{r:R18} is very efficient as almost all CO$^+$ forms HCO$^+$\footnote{This assumption may not be true if the UV radiation field is much higher than the usual ISM and the molecular fraction is { small. In such a case,} the destruction of CO$^+$ would involve atomic hydrogen and free electrons instead of H$_2$.}.

The destruction of HCO$^+$ is always dominated by free electrons:
\begin{align}
\rm HCO^+ + e \rightarrow  CO + H \tag{R21}.
\label{r:R21}
\end{align}

Thus, in chemical equilibrium, the HCO$^+$ abundance ($\rm {\it x}(HCO^+)$) can be written as:
\begin{align}
\rm {\it x}(HCO^+) =&\ \rm \frac{{\it x}(C^+)}{{\it x}(e)k_{R21}} \times [{\it x}(OH)k_{R15}+{\it x}(H_2O)k_{R19} \notag \\
&\ \rm + \frac{{\it x}(O)}{{\it x}(C^+)}{\it x}(CH)k_{R20}].
\label{eq:xhcop1}
\end{align}
{ By substituting equations \ref{eq:noh1}-\ref{eq:doh1} to the above and relating $\rm {\it x}(H_2O)$ with $\rm {\it x}(OH)$, we obtain:}
\begin{align}
\rm {\it x}(HCO^+) =&\ \rm \frac{{\it x}(C^+){\it x}(OH)}{{\it x}(e)k_{R21}} \times [k_{R15}+\epsilon k_{R19} \notag \\
&\ \rm + \frac{{\it x}(O)}{{\it x}(C^+)}\frac{{\it x}(CH)}{{\it x}(OH)}k_{R20}],
\label{eq:xhcop2}
\end{align}
where 
$\epsilon$ is:
\begin{align}
\rm \epsilon =&\ \rm [k_{pd}(R14)+{\it n}(H){\it x}(C^+)(k_{R15}+k_{R16})] \times  \notag \\
&\ \rm \{ k_{pd}(R12)+[k_{pd}(R12)+{\it n}(H){\it x}(C^+)k_{R13}] \times \notag \\
&\ \rm [\frac{k_{R9}+k_{R10}}{k_{R11}}+\frac{k_{R9}+k_{R10}+k_{R11}}{k_{R11}}\frac{{\it x}(e)k_{R6}}{{\it x}(H_2)k_{R5}}] \}^{-1}.
\end{align}

{ For low values of $\zeta_2$ (e.g., $\zeta_2$ $\lesssim$ 10$^{-15}$\,s$^{-1}$)}, { the production of electrons results primarily from the ionization of atomic carbon. The electron abundance can be, thus, approximated with the C$^+$ abundance}, e.g. $\rm {\it x}(e)$ = $\rm {\it x}(C^+)$ \citep{Goldsmith2001}. If the ionization degree of the cloud is large or the gas density is low, {\it x}(O)/x(C$^+$) $\approx$ 2, and the abundance of CH is an order of magnitude lower than that of OH. Thus, the contribution from the last term of equation \ref{eq:xhcop2} is small. Equation \ref{eq:xhcop2} can, then, be simplified as:
\begin{align}
\rm {\it x}(HCO^+) \approx &\ \rm \frac{{\it x}(OH)}{k_{R21}} \times [k_{R15}+\epsilon k_{R19}].
\label{eq:xhcop3}
\end{align}
{ For higher $\zeta_2$ (e.g., $\zeta_2$ $\gg$ 10$^{-15}$\,s$^{-1}$), a significant fraction of electrons are produced by H$^+$ \citep{LePetit2016}. In this case, equation \ref{eq:xhcop2} should be applied.} 

However, we should still keep in mind that at low extinctions and high FUV intensities, the destruction of CO$^+$ by H and free electrons becomes { more} efficient than that of H$_2$ (Reaction \ref{r:R18}). The formation of HCO$^+$ is, { therefore,} interrupted. The response of HCO$^+$ abundance does not follow a similar trend to that of OH at low $A_{\rm V}$ and high FUV intensities (as shown in Fig.~\ref{fig:pdr_den}). { In such a case, equation~\ref{eq:xhcop3} is not valid}.

\begin{figure}
\includegraphics[width=1.0\linewidth]{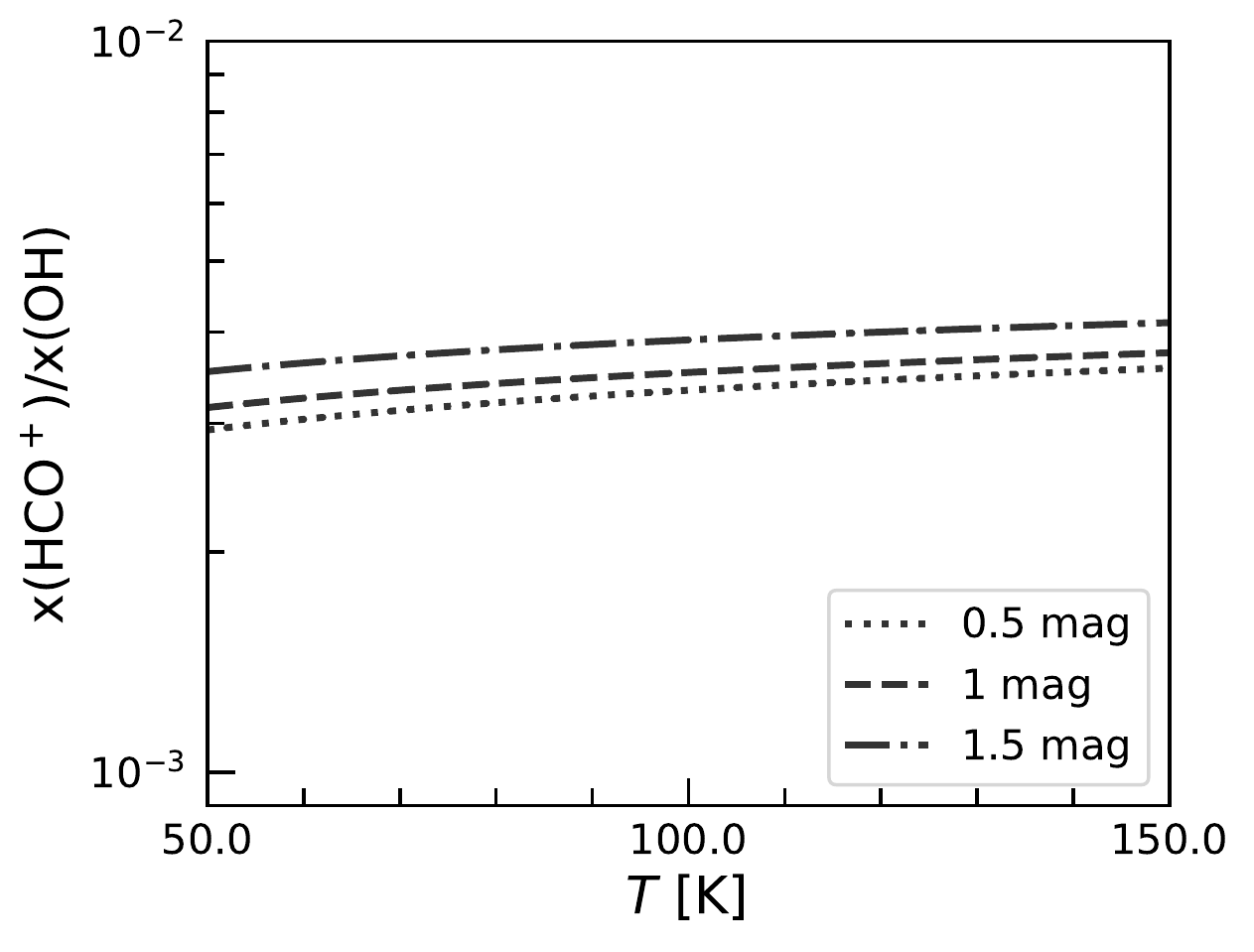}
\caption{The HCO$^+$/OH abundance ratios that calculated from equation \ref{eq:xhcop3} at a temperature range of 50--150\,K. Different line styles represent the three $A_{\rm V}$ values (0.5, 1.0, and 1.5\,mag) that are used in the calculation.}
\label{fig:rhcop_oh}
\end{figure}

{ Overall,} equation \ref{eq:xhcop3} shows that $x$(HCO$^+$) is proportional to {\it x}(OH) in diffuse clouds. High-sensitivity absorption observations found that the integrated optical depth of HCO$^+$ has a tight linear relation with that of OH \citep[$N$(HCO$^+$)/$N$(OH) = 0.03,][]{Liszt1996}. Considering a diffuse cloud with gas temperature ranging from 50 to 100\,K, at an extinction range of $0.5-1.5\,{\rm mag}$, the resultant HCO$^+$/OH ratio is nearly constant ($\sim$3.5$\times$10$^{-3}$, see Fig.~\ref{fig:rhcop_oh}). Note that while this value is an order of magnitude lower than the above observations, the abundance of OH measured in diffuse gas has large uncertainty between different LOSs (see Section \ref{sec:compare between obs and model} for more discussion). { Additionally,} most of the HCO$^+$ observations { of} \citet{Liszt1996} have high optical depths ($\tau >$1). { This means} that the formation and destruction of HCO$^+$ may be different from what we considered here.  
{ Furthermore, the CH abundance in translucent clouds is likely to be comparable to that of OH \citep[$\sim$4$\times$10$^{-8}$,][]{Liszt2002,Sheffer2008}, meaning that the HCO$^+$/OH ratio could be underestimated by equation \ref{eq:xhcop3}.}

{ Finally}, in the case of a high CO abundance, the latter becomes the precursor of HCO$^+$ through reaction:
\begin{align}
\rm CO + H_3^+ \rightarrow  HCO^+ + H_2 \tag{R22}.
\label{r:R22}
\end{align}
In this case, the abundance of HCO$^+$ may be underestimated.

\subsection{The abundance of CO in diffuse cloud}\label{sec:co abundance} 

There are three major formation paths of CO: Reactions \ref{r:R16}, \ref{r:R21}, and
\begin{align}
\rm CH + O \rightarrow &\ \rm CO + H \tag{R23}.
\label{r:R23}
\end{align}
The destruction of CO in the diffuse and translucent clouds is dominated by photodissociation:
\begin{align}
\rm CO + h\nu \rightarrow &\ \rm C + O \tag{R24},
\label{r:R24}
\end{align}
and by the interaction with He$^+$
\begin{align}
\rm CO + He^+ \rightarrow &\ \rm C^+ + O + He \tag{R25}.
\label{r:R25}
\end{align}

The CO abundance ($\rm {\it x}(CO)$) in chemical equilibrium can be written as:
\begin{align}
\rm {\it x}(CO) = &\ \rm \frac{{\it n}(H)}{k_{pd}(R24)+\eta_{CO}}[{\it x}(C^+){\it x}(OH)k_{R16}+\notag \\
&\ \rm {\it x}(HCO^+){\it x}(e)k_{R21}+{\it x}(CH){\it x}(O)k_{R23}].
\label{eq:xco1}
\end{align}
$\rm \eta_{CO}$ denotes the destruction rate of CO by He$^+$, which is proportional to the CRIR.
If we substitute {\it x}(HCO$^+$) as {\it x}(OH) using equation \ref{eq:xhcop2}, we obtain:
\begin{align}
\rm {\it x}(CO) = &\ \rm  \frac{{\it n}(H)}{k_{pd}(R24)+\eta_{CO}}[{\it x}(CH){\it x}(O)(k_{R20}+k_{R23})+ \notag \\
&\ \rm {\it x}(C^+){\it x}(OH)(k_{R15}+k_{R16}+\epsilon k_{R19})].
\label{eq:xco2}
\end{align}
The CO abundance is proportional to the gas density. In low-density diffuse gas where photodissociation dominates the destruction of CO, the OH channels dominate the formation of CO \citep{Luo2023}. Equation \ref{eq:xco2} can be simplified as :
\begin{align}
\rm {\it x}(CO) \approx &\ \rm  \frac{{\it n}(H){\it x}(C^+){\it x}(OH)}{k_{pd}(R24)}[(k_{R15}+k_{R16}+\epsilon k_{R19})].
\label{eq:xco3}
\end{align}
Since $\rm {\it x}(OH)$ is proportional to the CRIR, $\rm {\it x}(CO)$ increases with the increasing CRIR. 

\subsection{Comparison between the observed molecular abundances and model predictions}\label{sec:compare between obs and model} 

Radio emission line observations toward high latitude diffuse clouds (0.4$\lesssim$ $A_{\rm V}$ $\lesssim$1.1\,mag) have reported an abundance of OH between 1.6$\times$10$^{-7}$ and 4$\times$10$^{-6}$ \citep{Magnani1988}. Absorption measurements toward quasars at radio wavelengths suggest a { relative abundance ratio to total H column density} ($N_{\rm OH}$/$N_{\rm H}$) { of} $2.5-5\times10^{-8}$ at $A_{\rm V}$ of 1\,mag \citep{Crutcher1979,Liszt1996}. Considering that the molecular fraction is 10\%-20\% at such LOSs \citep{Lucas1996}, the abundance of OH { with respect to H$_2$} is in the range of $\sim$10$^{-7}$ to 10$^{-6}$. 
{ Recent radio observations by \citet{Tang2021a} covering a broad range of $A_{\rm V}$ (0.2$-$60\,mag)} suggest that the abundance of OH is higher ($\sim$10$^{-6}$) at low $A_{\rm V}$ and lower ($\sim$10$^{-7}$) at higher $A_{\rm V}$. The high spatial resolution ($\sim$0.12\,pc) observations by \citet{Xu2016a} in the Taurus boundary also found a decreasing trend of OH abundance with the $A_{\rm V}$.

Our models show that the abundance of OH strongly depends on the density distribution and the ISM environmental parameters (see Section \ref{sec:models}, Fig.~\ref{fig:pdr_den} and \ref{fig:pdr_cr}) and can vary by more than two orders of magnitude ($\sim$ 10$^{-9}$ $-$ 10$^{-7}$ at the same $A_{\rm V}$). As can be seen from equation \ref{eq:xoh2}, {\it x}(OH) is anti-correlated with ${\rm k_{pd}(R22)}$ and $n_{\rm H}$. This is consistent with the OH survey in nearby molecular clouds of \citet{Tang2021a}. Considering { the aforementioned large scatter of OH abundance in diffuse clouds}, we adopt an OH abundance in the range of 10$^{-7}$ to 10$^{-6}$ as the ``typical value". Thus, for any given FUV intensity (1$\leq$ $\chi$/$\chi_0$ $\leq$100) and density (10$^2$ $<$ $n_{\rm H}$ $\leq$ 10$^{3.5}$\,cm$^{-3}$), the model underestimates the abundance of OH at an $A_{\rm V}\simeq0.2-2\,{\rm mag}$ if a $\zeta_2 = 10^{-16}\,{\rm s}^{-1}$ is adopted (Fig.~\ref{fig:pdr_den}). As seen from Fig.~\ref{fig:pdr_cr}, the CRIR should be no less than 10$^{-16}$\,s$^{-1}$ if we are to reproduce the observed abundance of OH in diffuse clouds.

Due to the sub-thermal excitation of HCO$^+$ transitions in low-density gas \citep{Godard2010,Luo2020}, its abundance can vary over an order of magnitude without precise measurement of both the excitation temperature and optical depth \citep{Luo2023}. The abundance of HCO$^+$ in diffuse clouds has been measured frequently through absorption observations against strong continuum sources (e.g., quasars, H\,{\sc ii} regions). Despite the different methods in obtaining the column density of H$_2$, the abundance of HCO$^+$ is fairly constant in diffuse gas \citep[(1.7--3.1)$\times$10$^{-9}$,][]{Lucas1996,Liszt2016,Gerin2019,Luo2020}. At a CRIR of 10$^{-16}$\,s$^{-1}$, our models underestimate the abundance of HCO$^+$ in diffuse gas in all density and FUV intensity ranges explored (Fig.~\ref{fig:pdr_den}). Therefore, to reproduce the observed abundance of HCO$^+$, the gas density should be approximately in the range of 10$^2$ $\leq$ $n_{\rm H}$ $\leq$ 10$^{2.5}$\,cm$^{-3}$ and the CRIR should be $\zeta_2 > 10^{-16}\,{\rm s}^{-1}$ (Fig.~\ref{fig:pdr_cr}). The inferred density is consistent with the inferred density in Section \ref{sec:nh} and those by radio and UV absorptions \citep[e.g., 80$-$160\,cm$^{-3}$,][]{Goldsmith2013,Liszt2016,Luo2020}.

In diffuse clouds, most of the carbon is { in the form of} C$^+$ or C$^0$ due to insufficient shielding from UV photons. Similar to HCO$^+$, the CO low-$J$ transitions are usually sub-thermally excited in the low-density diffuse gas \citep{Goldsmith2008,Luo2020}. The abundance of CO in diffuse clouds can vary by two orders of magnitude in different environments (e.g., 2.6$\times$10$^{-8}$--2$\times$10$^{-5}$ from UV absorption measurements). { In particular, it} increases with increasing $N_{\rm H_2}$ (or $A_{\rm V}$) \citep{Burgh2007,Liszt2007,Sheffer2008}. The abundance of CO through CO (J=1--0) emission line measurements in low extinction regions ($A_{\rm V}\sim1\,{\rm mag}$) in Taurus\footnote{The value is obtained by averaging the pixels without CO detection.} is approximately (1.2--7)$\times$10$^{-6}$ \citep{Goldsmith2008,Pineda2010}, which is over an order of magnitude lower than the canonical value in well-shielded regions \citep[e.g., $\sim$10$^{-4}$,][]{Frerking1982}. This value is similar to that of absorption observations in diffuse LOSs \citep[$A_{\rm V}$ = 0.19--2.08\,mag, $f_{\rm CO}$ = (0.2$\pm$0.1--5$\pm$4)$\times$10$^{-6}$,][]{Luo2020}. Our models shown in Fig.~\ref{fig:pdr_cr} can reasonably explain the observed CO abundances.  

Considering all the above, we find that the abundances of OH, HCO$^+$, and CO in diffuse clouds suggest an ISM environment with low gas densities ($n_{\rm H}$ $\lesssim$10$^{2.5}$\,cm$^{-3}$) and high CRIRs ($\zeta_2$ $\gtrsim$10$^{-16}$\,s$^{-1}$). A more quantitative analysis is discussed below (\S\ref{sec:CRIR}).

\section{The abundance ratios and CRIR in diffuse clouds}\label{sec:abundance ratio}

As can be seen from equations \ref{eq:xoh2} and \ref{eq:xhcop3}, the abundances of OH and HCO$^+$ are proportional to the CRIR in diffuse gas, which implies they can be used to constrain CRIR with a given density. However, molecular hydrogen does not emit radiation that can be observed from radio telescopes due to the lack of permanent dipole moment, { the measurement of an accurate H$_2$ column density in the CNM, as well as its exact abundance, is very hard}. 
Instead, it is possible to put constraints on the CRIR { by combining the} observed molecular column densities { and their ratios with chemical models}.

\subsection{The abundance ratio of OH/CO}\label{sec:abundance ratio of oh2co}

{ The abundance ratio of OH/CO can be obtained} from equation \ref{eq:xco2}:
\begin{align}
\rm \frac{{\it x}(OH)}{{\it x}(CO)} = &\ \rm  \frac{1}{{\it x}(C^+)(k_{R15}+k_{R16}+\epsilon k_{R19})} \times \notag \\
&\ \rm [\frac{k_{pd}(R24)+\eta_{CO}}{{\it n}(H)}-\frac{{\it x}(CH)}{{\it x}(CO)}{\it x}(O)(k_{R20}+k_{R23})].
\label{eq:rohco1}
\end{align}
{ The second term in the square brackets can be safely ignored in the diffuse gas\footnote{The term ${\rm k_{pd}(R24)}$+${\rm \eta_{CO}}$ is comparable to ${\rm k_{R20}+k_{R23}}$ at $A_{\rm V}$ = 1\,mag, while 1/${\rm {\it n}(H)}$ is apparently a few orders of magnitude higher than ${\rm \frac{{\it x}(CH)}{{\it x}(CO)}{\it x}(O)}$, thus, the second term can be ignored.}}:
\begin{align}
\rm \frac{{\it x}(OH)}{{\it x}(CO)} \approx &\ \rm  \frac{k_{pd}(R24)+\eta_{CO}}{{\it n}(H){\it x}(C^+)(k_{R15}+k_{R16}+\epsilon k_{R19})}.
\label{eq:rohco2}
\end{align}

\begin{figure*}
\includegraphics[width=1.0\linewidth]{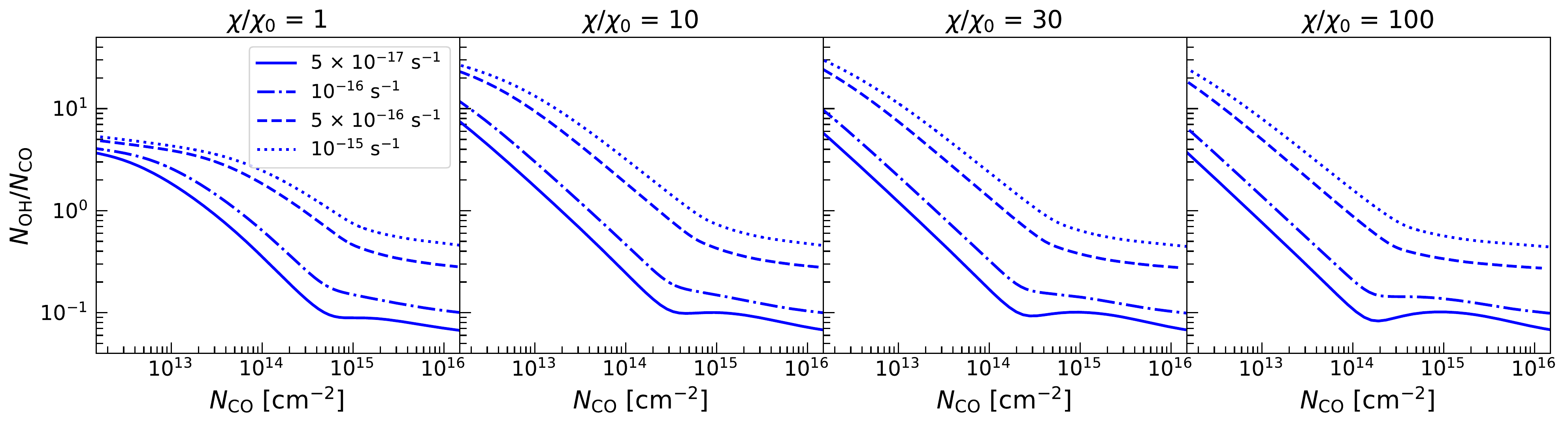}
\caption{The predicted abundance ratios by PDR models with $n_{\rm H}$ = 10$^2$\,cm$^{-3}$, and different line styles represent the four CRIR values that are used for modelling ($\rm \zeta_2$ = 5$\times$10$^{-17}$, 10$^{-16}$, 5$\times$10$^{-16}$, and 10$^{-15}$\,s$^{-1}$). From left to right represent the four representative FUV intensities ($\chi$/$\chi_0$ = 1, 10, 30, and 100).}
\label{fig:r_oh_co_1d}
\end{figure*}

Figure \ref{fig:r_oh_co_1d} shows the predicted abundance ratio of OH/CO with the column density of CO { ($N_{\rm CO}$)} from 1D slab model simulations ($n_{\rm H}$ = 10$^2$\,cm$^{-3}$). The abundance ratio of OH/CO decreases as the increasing $N_{\rm CO}$.
The abundance ratio of OH/CO will increase { with the increasing FUV intensity} by a factor of a few at low $N_{\rm CO}$, { while it remains approximately constant at high $N_{\rm CO}$}. This is because, at a low FUV intensity (e.g., $\chi$/$\chi_0\sim1$), the gas temperature is significantly lower than that of a moderate FUV intensity (e.g., $\chi$/$\chi_0\sim10$). The decreasing gas temperature would increase the reaction rates of \ref{r:R15}, \ref{r:R16}, and \ref{r:R19}, leading to a lower OH/CO ratio. { However, at high $N_{\rm CO}$, CO molecules exist mainly in well-shielded regions, in which the gas temperature does not significantly increase even when high external FUV intensities exist.} 

As shown in Fig.~\ref{fig:r_oh_co_1d}, the abundance ratio of OH/CO monotonically increases with the increasing CRIR at a given $N_{\rm CO}$. { Since high CRIR heats the gas,} the reaction rates of \ref{r:R15}, \ref{r:R16}, and \ref{r:R19} will decrease. { At the meantime, $\eta_{\rm CO}$ will increase as the increasing CRIR,} leading to a higher OH/CO ratio. 

The results shown in Fig.~\ref{fig:r_oh_co_1d} indicate that once the gas density can be constrained from line ratios (e.g., C\,{\sc i}/CO, CO rotational line ladder), the CRIR can be inferred from OH and CO observations without knowledge of their exact abundance relative to H$_2$. This { is potentially useful} in high spatial resolution radio spectral line observations, { especially when the column densities of H$_2$ cannot be determined.} 

However, the main challenge of generalizing the { use of OH and CO as a probe of CRIR} is that the excitation temperature of OH is usually within a few K { above} the Galactic synchrotron background \citep{Li2018}. { This will lead to an order of magnitude uncertainty in $N_{\rm OH}$ through emission lines if $T_{\rm ex}$ varies between 0.1--1.0\,K above $T_{\rm bg}$ (note that $N_{\rm OH}$ $\propto$ $T_{\rm ex}$/($T_{\rm ex}$-$T_{\rm bg}$)).} Absorption measurements { are less suffered from the above issues ($N_{\rm OH}$ $\propto$ $T_{\rm ex}$)}, while the optical depth of OH is usually between a few to tens of 10$^{-3}$ (more than an order of magnitude lower than that of HCO$^+$). { This affects} the feasibility of detecting the absorption of OH in a relatively short integration time even with the JVLA. With the proposed capability of high-sensitivity radio telescopes such as FAST, and SKA in the future \citep{McClure-Griffiths2015}, it is possible to constrain the CRIR through OH and CO in both the Milky Way and external galaxies.

\subsection{The abundance ratio of HCO$^+$/CO}\label{sec:abundance ratio of hcop2co}

{ Contrary to OH,} HCO$^+$ can be easily detected in diffuse clouds through absorption measurements against strong continuum sources (e.g., a few to tens of minutes with ALMA), { and has less uncertainty in calculating its column density}. 

Following the same way, { if we replace equation \ref{eq:xhcop3} with equation \ref{eq:rohco2}, the abundance ratio of HCO$^+$/CO can be written as}:
\begin{align}
\rm \frac{{\it x}(HCO^+)}{{\it x}(CO)} \approx &\ \rm  \frac{k_{pd}(R24)+\eta_{CO}}{{\it n}(H){\it x}(C^+)k_{R21}} [1+\frac{k_{R16}}{k_{R15}+\epsilon k_{R19}}]^{-1}.
\label{eq:rhcopco1}
\end{align}

\begin{figure*}
\includegraphics[width=1.0\linewidth]{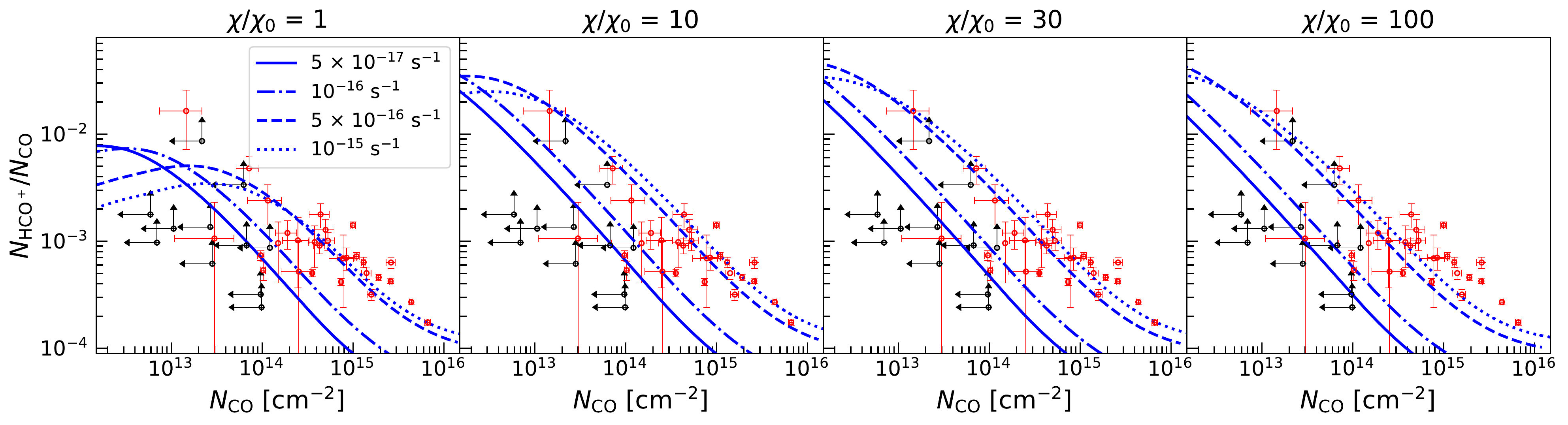}
\caption{The observed relative abundance ratio (column density ratio) of $N_\mathrm{HCO^+}$/$N_\mathrm{CO}$ in \citet{Luo2020} overlaid on the predicted abundance ratios by PDR models { with $n_{\rm H}$ = 10$^2$\,cm$^{-3}$. From left to right represent} the four representative FUV intensities ($\chi$/$\chi_0$ = 1, 10, 30, and 100). Red dots denote observations with both HCO$^+$ and CO absorption detections. Black dots denote HCO$^+$ absorptions without CO detection, it is thus a lower limit for $N_\mathrm{HCO^+}$/$N_\mathrm{CO}$. The four CRIR values used for modelling are 5$\times$10$^{-17}$, 10$^{-16}$, 5$\times$10$^{-16}$, and 10$^{-15}$\,s$^{-1}$.}
\label{fig:r_hcop_co_1d}
\end{figure*}

Figure \ref{fig:r_hcop_co_1d} shows the predicted abundance ratio of HCO$^+$/CO with $N_{\rm CO}$ from 1D slab model simulations { ($n_{\rm H}$ = 10$^2$\,cm$^{-3}$)}, overlaid with the observed values from absorption measurements. Red dots are the observed values, and different line curves represent PDR models under different CRIRs. As seen from Fig.~\ref{fig:r_hcop_co_1d}, the majority (22 out of the total 26) of the observational values are within $10^{-16}\lesssim\zeta_2\lesssim10^{-15}\,{\rm s}^{-1}$. 

At a given $N_{\rm CO}$, the abundance ratio of HCO$^+$/CO will first increase with increasing CRIR. This is similar to the variance of OH/CO as we have explained in Section \ref{sec:abundance ratio of oh2co}. With the increasing photodissociation rate (\ref{r:R24}) and the gas temperatures, reaction rates \ref{r:R15}, \ref{r:R16}, \ref{r:R19}, and \ref{r:R23} decrease while the reaction rates of electron recombination (\ref{r:R21}) increase. Thus, the increasing trend of HCO$^+$/CO is slowing down because of the increasing destruction rate of HCO$^+$ by free electrons. { Consequently}, there is a plateau of the HCO$^+$/CO ratio (as well as the abundance of HCO$^+$), where the increasing CRIR no longer increases the HCO$^+$/CO ratio. 

However, the HCO$^+$/CO ratio decreases at an even higher CRIR. This is because the most important precursors of HCO$^+$ (\ref{r:R18}), { such as} CO$^+$, are destroyed by atomic hydrogen and free electrons at high temperature, reducing the formation efficiency of HCO$^+$. The abundance ratio of HCO$^+$/CO reaches a maximum when the CRIR is approximately 10$^{-15}$\,s$^{-1}$ for $n_{\rm H}=10^2\,{\rm cm}^{-3}$.

\subsection{The CRIR inferred from HCO$^+$ and CO absorptions}\label{sec:CRIR}

In order to quantitatively constrain $\zeta_2$ toward each source, we have run PDR models with { ${\rm log_{10}\, (\zeta_2/ s^{-1})}=-18$ to $-14$ with a step size of 0.07~dex and ${\rm log_{10}\, ({\it n}_{\rm H}/ cm^{-3})}$ = 1 -- 3.5 with a step size of 0.1~dex, interacting with three FUV intensities ($\chi/\chi_0=1,5,10$). We vary $\zeta_2$ and $n_{\rm H}$} to find the optimum model by minimizing the reduced $\chi^2$ function:
\begin{equation}
\rm \chi^2_{red} = \frac{1}{N}\sum_i  \frac{\left ( f_{obs}^i - f_{model}^i \right )^2}{{\sigma^i_{obs}}^2},
\end{equation}
where $\rm {f_{obs}^i}$ and ${\rm \sigma^i_{obs}}$ are the observed molecular column densities and their corresponding uncertainties of the $i$th species, respectively. $\rm {f_{model}^i}$ is the column density from PDR models and N = 1 is the degree of freedom. 

\begin{figure*}
\includegraphics[width=1.0\linewidth]{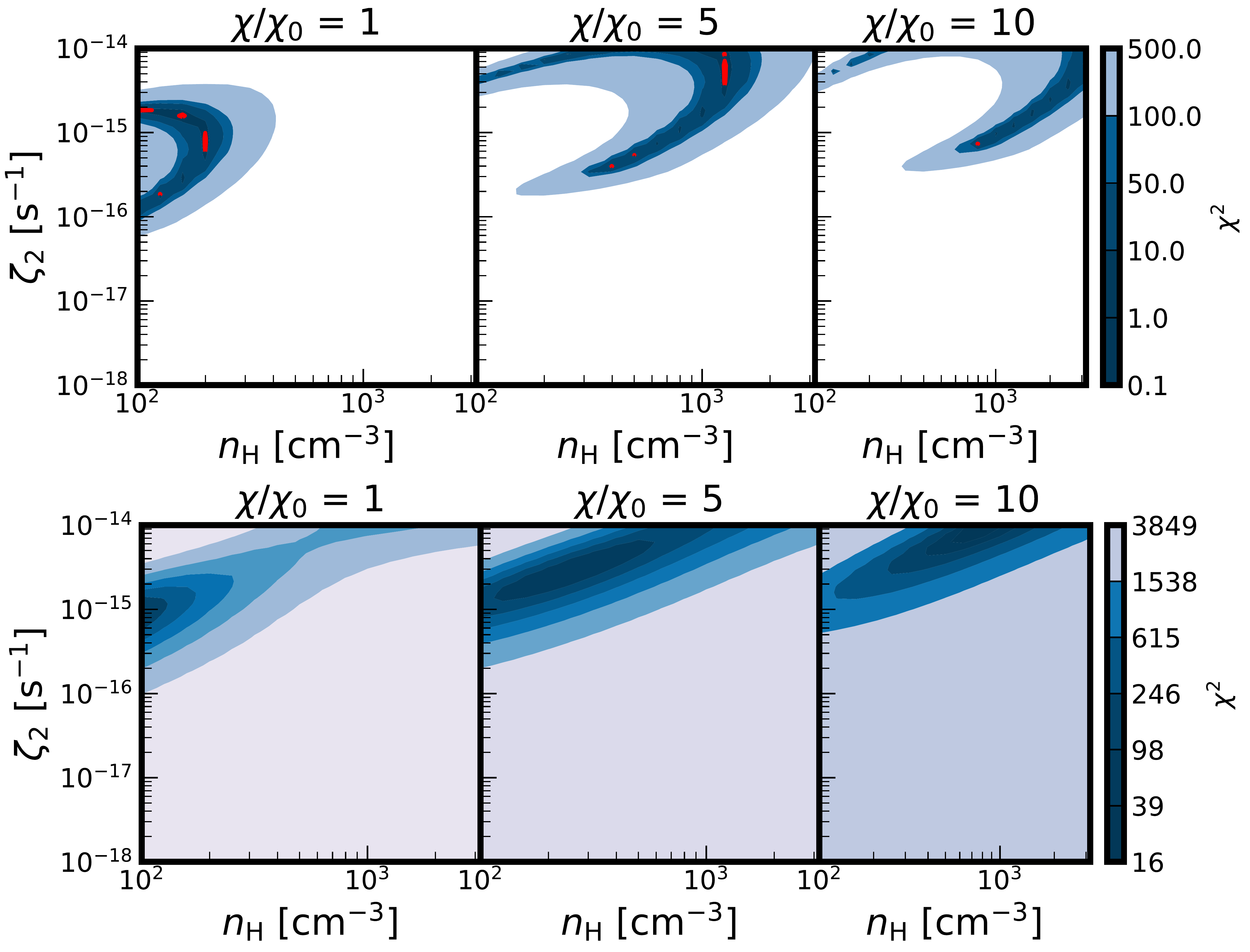}
\caption{The ${\rm \chi^2_{red}}$ distribution toward 3C454.3 V1 (up) and J0325+2224 (low) in the given parameter space. From left to right: ${\rm \chi^2_{red}}$ distribution at $\chi/\chi_0$ =1, 5, and 10. Red contours represent the boundary of where the deviation between modelled values and observations is 1$\sigma$.}
\label{fig:chi2}
\end{figure*}

{ We show two representative ${\rm \chi^2_{red}}$ distributions from our fitting in Figure \ref{fig:chi2}.} 
The higher values of ${\rm \chi^2_{red}}$ are colored with bright blue and the lower colored with dark blue. { White color indicates values beyond the maximum in this colour-bar.} Red contours represent the boundary of the best-fit parameter where the deviation between modeled values and observations is 1$\sigma$. { We treat a fitting result with a minimal value of ${\rm \chi^2_{red}} \lesssim 1$ (upper panel of Figure \ref{fig:chi2}) as a ``high confidence'' fit, and a fitting result with a minimal value of ${\rm \chi^2_{red}} \gg 1$ (lower panel of Figure \ref{fig:chi2}) as a ``low confidence'' fit. 

As seen from Figure \ref{fig:chi2}, the best-fit values drift to both higher $\zeta_2$ and $n_{\rm H}$ as the increasing FUV intensity. 
However, at high FUV intensity, the best-fit value of $n_{\rm H}$ from PDR models is much higher than the allowable density range by MCMC runs (Table \ref{tab:nh}. On the other hand, increasing $\chi$/$\chi_0$ would increase $T_{\rm k}$. In that case, the inferred gas density should be lower (as seen in Figure \ref{fig:tk_nh}). Thus, it is not likely that FUV intensity is high for our sources. 

Here, we made a rough estimation of the FUV intensity in diffuse clouds. The interstellar radiation field is a function of dust temperature \citep[$T_{\rm d}$,][]{Beuther2014}:
\begin{equation}
    \chi/\chi_0 = \frac{1}{1.7}\frac{4.7\times10^{-31}}{5.6\times10^{-26}}\frac{1}{\gamma(A_{\rm V})}T_{\rm d}^6,
\end{equation}
where $\gamma(A_{\rm V})$ is the attenuation factor. We consider $\gamma(A_{\rm V})$ = 0.35 at $A_{\rm V}$ = 1\,mag \citep{Glover2012}. For $T_{\rm d}$ = 15\,K, the derived FUV intensity is $\chi/\chi_0$ = 1.4. A variation on $T_{\rm d}$ of $\pm$2\,K would only result in a difference on $\chi/\chi_0$ by a factor of 2. In the following, we only consider the fitting results at $\chi$/$\chi_0$ = 1.}

The fitting results of all sources are summarised in Table \ref{tab:col}, { in which high confidence fit parameters are labelled with a ``*'' at the end of each row. The $\zeta_2$ in our sample (high confidence results) is in the range of $\zeta_2\sim1.0_{-1.0}^{+14.8}$ $\times$10$^{-16}- 2.5_{-2.4}^{+1.4}$ $\times$10$^{-15}$\,s$^{-1}$. The $n_{\rm H}$ is in the range of (0.1--3.2)$\times$10$^2$\,cm$^{-3}$, which is consistent with that obtained from MCMC runs in Section \ref{sec:nh}.}

\begin{figure*}
\includegraphics[width=1.0\linewidth]{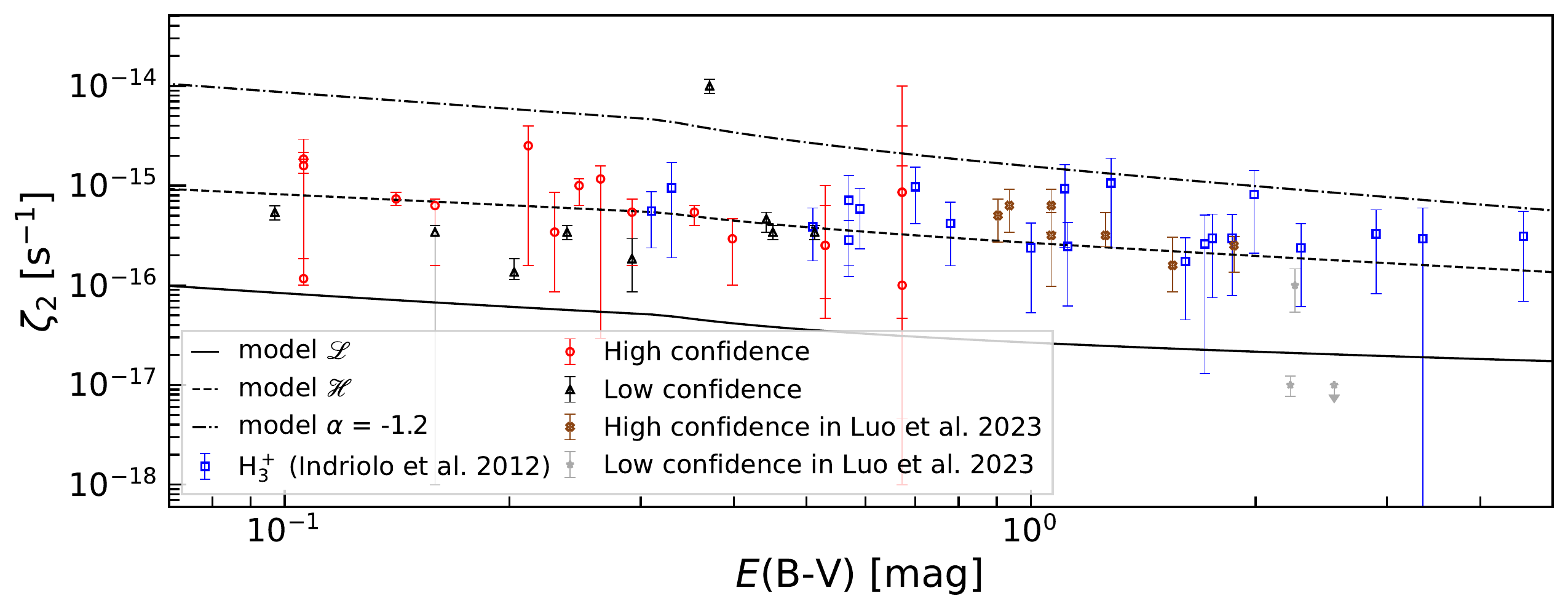}
\caption{{ The red circles (high confidence fit) and black triangles (low confidence fit) show the inferred $\zeta_2$} in this work. Blue squares denote the measurements from H$^+_3$ by \citet{Indriolo2012}. Brown crosses and gray stars are the high confidence fit and low confidence fit of $\zeta_2$ toward nearby star-forming cloud--IC~348. Black dashed, black solid, and black dash-dotted curves represent the polynomial fit to $\zeta_2$ at different $E$(B-V) for models $\mathscr{L}$, $\mathscr{H}$ and with low-energy spectral slope $\alpha$ = $-$1.2 by \citet{Padovani2022}.}
\label{fig:av_cr}
\end{figure*}

{ In Fig.~\ref{fig:av_cr}, we plot the \citet{Luo2023} $\zeta_2$ toward IC~348 measured from HCO$^+$ and CO measurements, and the \citet{Indriolo2012} $\zeta_2$ from H$_3^+$ measurements. For comparison, we also include} the \citet{Padovani2022} CR attenuation models $\mathscr{L}$, $\mathscr{H}$, and with low-energy spectral slope = $-$1.2. To plot $\zeta_2$ as a function of $E$(B-V), we convert N$_{\rm H_2}$ to $E$(B-V) by assuming that the atomic fraction $f_{\rm atomic}$ follows the power law: $f_{\rm atomic}$ = 0.64/$A_{\rm V}$ at $A_{\rm V} >$1 mag \citep{Luo2023}, and the $f_{\rm atomic}$ = 0.64\footnote{The atomic fraction is between 0.6-0.8 at low extinction LOSs \citep{Luo2020}.} at $A_{\rm V} \leq $1 mag\footnote{Note that the discontinuity seen in models at $E$(B-V) = 0.32 mag is not a peculiarity of the models themselves, but it is due to the assumption on $f_{\rm atomic}$.}.
The average value of CRIR in our samples (red dots) is (8.0$\pm$6.4)$\times$10$^{-16}$\,s$^{-1}$. This is $\sim2$ times higher than the value measured with H$^+_3$ toward nearby massive stars ($\zeta_2$ = 3.5$^{+5.3}_{-3.0}$ $\times$10$^{-16}$\,s$^{-1}$) and toward IC~348 (4.7$\pm$1.5\,$\times$10$^{-16}$\,s$^{-1}$, see  Fig.~\ref{fig:av_cr}). Considering that our sightlines are located at much lower $E$(B-V)\footnote{We convert $A_{\rm V}$ to $E$(B-V) by adopting $A_{\rm V}$ = 3.1$E$(B-V) \citep{Schlafly2011}.} { (0.1 -- 0.7\,mag), the values are reasonably consistent with the low $E$(B-V) portion} by \citet{Indriolo2012}. { Our results are consistent with the model $\mathscr{H}$ by \citet{Padovani2018}, in which CRs (and consequently the CRIR) are attenuated as a function of $A_{\rm V}$} \citep{Padovani2018, Padovani2022, Gaches2022}. However, model $\mathscr{L}$, which corresponds to a ``low'' cosmic-ray spectrum based on Voyager-1 data, is found to underestimate the CRIR by almost an order of magnitude.

\section{Conclusions}\label{sec:conclusion}

{ We analyze the abundances of CO, OH, and HCO$^+$, and their abundance ratios in chemical equilibrium. We present a new approach to constrain the CRIR using these species. We calculated the column densities of HCO$^+$ and CO toward diffuse LOSs against quasars and compare them with {\sc 3d-pdr} models. Our inferred values of $\zeta_2$ show good consistency with previous measurements using H$^+_3$.} The main conclusions are as follows:

\begin{enumerate}
    \item { The gas volume densities ($n_{\rm H_2}$) obtained from CO (1--0) and (2--1) transitions toward 4 LOSs are in the range of (0.14$\pm$0.03 -- 1.2$\pm$0.1)$\times$10$^2$\,cm$^{-3}$ (at $T_{\rm k}$ = 50\,K), which suggests a sub-thermal excitation environment for CO and HCO$^+$.}

    \item  { Analyzing the chemical response of different molecules, we found that} the abundance of CO increases with gas density and decreases with increasing FUV intensity, while the abundances of OH and HCO$^+$ mostly have an opposite trend to that of CO.
    
    \item  Our { analytic expressions} give an excellent abundance of OH when $\zeta_2$ $<$10$^{-15}$\,s$^{-1}$. { This is potentially useful for hydrodynamical simulations as it reduces the computational expense of chemical networks}. In the diffuse gas, the abundance of OH is proportional to the CRIR. 

    \item  At a given $N_{\rm CO}$, the abundance ratio of OH/CO is monotonically increasing with the increase $\zeta_2$ in the diffuse cloud, while the abundance ratio of HCO$^+$/CO increases reaching a local maximum value at $\zeta_2$ $\approx$ 10$^{-15}$\,s$^{-1}$, before it decreases again. The downward trend of HCO$^+$/CO at high $\zeta_2$ is caused due to the increased destruction of the precursor of HCO$^+$--CO$^+$ by electrons and atomic H at higher gas temperatures.
    
    \item By comparing the observational values of HCO$^+$/CO and chemical models, we find that the average $\zeta_2$ in our sample is { (8.0$\pm$6.4)$\times$10$^{-15}$\,s$^{-1}$}. This value is $\sim$2 times higher than that of higher extinction regions, which is consistent with the hypothesis of decreasing $\zeta_2$ as the increasing $A_{\rm V}$ in theoretical studies.

\end{enumerate}

{ With high-sensitivity measurements from HCO$^+$ and CO, we have demonstrated the possibility of using HCO$^+$/CO ratios to constrain the CRIR in diffuse gas without knowing the exact molecular abundances relative to H$_2$.
We propose that the abundance ratios of OH/CO and HCO$^+$/CO can be used to constrain the CRIR, especially with interferometry observations where high-resolution H$_2$ information is inaccessible. 
Due to the difficulties in obtaining accurate excitation and optical depth of OH with current facilities, we foresee that future instruments, such as SKA, will produce large samples of data sets for which our approach will be potentially useful.}

\begin{acknowledgments}

We thank Marco Padovani for the useful discussions on the CR attenuation models and for providing the latest data of these models. 
This work has been supported by the National Natural Science Foundation of China (grant No. 12041305), China Postdoctoral Science Foundation (grant No. 2021M691533), the Program for Innovative Talents, Entrepreneur in Jiangsu, the science research grants from the China Manned Space Project with NO.CMS-CSST-2021-A08. P.~Z. acknowledges support from the National Natural Science Foundation of China (grant No. 12273010).

This paper makes use of the following ALMA data: ADS/JAO.ALMA$\#$2015.1.00503.S. ALMA is a partnership of ESO (representing its member states), NSF (USA) and NINS (Japan), together with NRC (Canada) and NSC and ASIAA (Taiwan) and KASI (Republic of Korea), in cooperation with the Republic of Chile. The Joint ALMA Observatory is operated by ESO, AUI/NRAO and NAOJ.

\end{acknowledgments}

%

\vspace{5mm}
\facilities{ALMA}


\software{CASA \citep{McMullin2007},
          Astropy \citep{Astropy2013,Astropy2018},  
          {\sc radex} \citep{Van2007},
          {\it emcee} \citep{Foreman2013},
          {\sc 3d-pdr} \citep{Bisbas2012}
          }

\appendix

\section{Derivation of abundance of OH in chemical equilibrium}\label{sec:xco}

The formation of OH is given by:
\begin{align}
\rm F[{\it n}(OH)] =&\  \rm {\it n}(H_2O^+){\it n}(e)k_{R6} \notag\\
&\ + \rm {\it n}(H_2O^+){\it n}(e)(k_{R9}+k_{R10}) \notag\\
&\ + \rm  {\it n}(H_2O)k_{pd}(R12),
\label{eq:noh1}
\end{align}
where $\rm {\it n}(H_2O^+)$ can be written as:
\begin{align}
\rm [{\it n}(H_2O^+)] =&\  \rm \frac{{\it n}(H_2O^+){\it n}(H_2)k_{R5}}{{\it n}(e)(k_{R9}+k_{R10}+k_{R11})},
\end{align}
and $\rm {\it n}(H_2O^+)$ can be written as:
\begin{align}
\rm [{\it n}(H_2O^+)] =&\  \rm \frac{{\it n}(OH^+){\it n}(H_2)k_{R4}}{{\it n}(H_2)k_{R5}+{\it n}(e)(k_{R6}+k_{R7}+k_{R8})}.
\end{align}
The destruction of H$_2$O is dominated by photodissociation and C$^+$ in the diffuse cloud (note that the destruction of H$_2$O will be dominated by H$^+$ in dense cloud), thus, we have
\begin{align}
\rm {\it n}(H_2O) =&\  \rm \frac{{\it n}(H_2O^+){\it n}(e)k_{R11}}{k_{pd}(R12)+{\it n}(C^+)k_{R13}}.
\end{align}
Then, equation \ref{eq:noh1} can be written as:
\begin{align}
\rm F[{\it n}(OH)] =&\  \rm \frac{{\it n}(OH^+){\it n}(H_2)k_{R4}}{{\it n}(H_2)k_{R5}+{\it n}(e)(k_{R6}+k_{R7}+k_{R8})} \notag\\
&\ \times \rm [{\it n}(e)k_{R6}+{\it n}(H_2)k_{R5}(1-\delta)],
\label{eq:noh2}
\end{align}
where $\delta$ is:
\begin{align}
\rm \delta =&\  \rm \frac{k_{R11}}{k_{R9}+k_{R10}+k_{R11}} \times \notag\\
&\ \rm \frac{{\it n}(C^+)k_{R13}}{k_{pd}(R12)+{\it n}(C^+)k_{R13}}.
\label{eq:delta}
\end{align}


Considering chemical equilibrium of $\rm {\it n}(OH^+)$, the term $\rm {\it n}(OH^+){\it n}(H_2)k_{R4}$ can be written as:
\begin{align}
\rm {\it n}(OH^+){\it n}(H_2)k_{R4} =&\  \rm {\it n}(O^+){\it n}(H_2)k_{R2}+{\it n}(O){\it n}(H_3^+)k_{R3} \notag \\
= &\ \rm {\it n}(O)[{\it n}(H^+)k_{R1} + \zeta_O + {\it n}(H_3^+)k_{R3}],
\label{eq:ohp1}
\end{align}
where $\zeta_{\rm O}$ $\approx$ 2.83$\zeta_2$ is the cosmic-ray ionization rate of atomic oxygen. 

H$^+$ is produced by cosmic-rays (CRs):
\begin{align}
\rm H + CRs  \rightarrow &\ \rm H^+ + e \tag{RA1}, \label{r:RA1}\\ 
\rm H_2 + CRs  \rightarrow &\ \rm H^+ + H + e \tag{RA2}, \label{r:RA2}\\ 
\rm H_2^+ + H \rightarrow &\ \rm H_2 + H^+ \tag{RA3}, \label{r:RA3}
\end{align}
and removed by atomic oxygen (reaction \ref{r:R1}) and electron recombination reaction:
\begin{align}
\rm H^+ + e \rightarrow H + h\nu \tag{RA4}.
\label{r:RA4}
\end{align}
In chemical equilibrium, $\rm {{\it n}(H^+)}$ is in the form of:
\begin{align}
\rm {\it n}(H^+) =&\  \rm \frac{{\it n}(HI)k_{RA1}+{\it n}(H_2)k_{RA2}+{\it n}(H_2^+)k_{RA3}}{{\it n}(e)k_{RA4}+{\it n}(O)k_{R1}},
\label{eq:hp1}
\end{align}
in which H$_2^+$ is formed through:
\begin{align}
\rm H_2 + CRs  \rightarrow &\  \rm H_2^+ + e \tag{RA5},\label{r:RA5}
\end{align}
and removed by H through reaction \ref{r:RA3} or by H$_2$ through:
\begin{align}
\rm H_2^+ + H_2 \rightarrow H_3^+ + H \tag{RA6}.
\label{r:RA6}
\end{align}
Thus, equation \ref{eq:hp1} can be written as:
\begin{align}
\rm {\it n}(H^+) =&\  \rm \frac{{\it n}(HI)k_{RA1}+{\it n}(H_2)k_{RA2}+{\it n}(H_2)k_{RA5}\xi}{{\it n}(e)k_{RA4}+{\it n}(O)k_{R1}},
\label{eq:hp2}
\end{align}
where $\rm {k_{RA1}}$ = $\zeta_1$ = 1/2$\zeta_2$, $\rm {k_{RA2}}$ = 0.02 $\zeta_2$, $\rm {k_{RA5}}$ = 0.88 $\zeta_2$, and $\xi$ is in the form of:
\begin{align}
\rm \xi =&\  \rm \frac{{\it n}(HI)k_{RA3}}{{\it n}(HI)k_{RA3}+{\it n}(H_2)k_{RA6}} \notag \\
= &\ \rm \frac{2-2{\it f}_{mol}}{2+1.25{\it f}_{mol}}.
\label{eq:xi}
\end{align}
Thus, if $\rm {{\it f}_{mol}}$ is close to 1, $\xi$=0. Otherwise, if $\rm {{\it f}_{mol}}$ is close to 0, $\xi$=1.

However, equation \ref{eq:hp2} does not consider the formation of H$^+$ by He$^+$. The formation of He$^+$ is mainly due to the He ionization due to CRs:
\begin{align}
\rm He + CRs \rightarrow &\ \rm He^+ + e \tag{RA7}.
\label{r:RA7}
\end{align}
The destruction of He$^+$ is either dominated by H or H$_2$ in diffuse cloud:
\begin{align}
\rm He^+ + H \rightarrow  H^+ + He \tag{RA8}, \label{r:RA8}\\
\rm He^+ + H_2 \rightarrow  H^+ + H + He \tag{RA9}, \label{r:RA9}\\
\rm He^+ + H_2 \rightarrow  H_2^+ + He \tag{RA10}.
\label{r:RA10}
\end{align}
In equilibrium, n(He$^+$) can be written as:
\begin{align}
\rm n(He^+) =&\  \rm \frac{{\it n}(He)k_{RA7}}{{\it n}(HI)k_{RA8}+{\it n}(H_2)(k_{RA9}+k_{RA10})},
\label{eq:hep}
\end{align}
where $\rm {k_{RA7}}$ = 0.5 $\zeta_2$.

If we take reactions \ref{r:RA8} and \ref{r:RA9} into account, equation \ref{eq:hp2} can be rewritten as:
\begin{align}
&\ \rm {\it n}(H^+) = \notag \\
&\  \rm \frac{{\it n}(HI)k_{RA1}+{\it n}(H_2)k_{RA2}+{\it n}(H_2)k_{RA5}\xi+{\it n}(He)k_{RA7}}{{\it n}(e)k_{RA4}+{\it n}(O)k_{R1}} \notag \\
= &\ \rm \frac{{\it n}(HI)k_{RA1}+{\it n}(H_2)k_{RA2}+{\it n}(H_2)k_{RA5}\xi+0.1{\it n}(H)k_{RA7}}{{\it n}(e)k_{RA4}+{\it n}(O)k_{R1}} \notag \\
= &\ \rm \frac{0.5[1.1-(0.98-0.88\xi){\it f}_{mol}]{\it n}(H)\zeta_2}{{\it n}(e)k_{RA4}+{\it n}(O)k_{R1}}.
\label{eq:hp3}
\end{align}

H$_3^+$ is formed through reaction \ref{r:RA6} and removed by electrons:
\begin{align}
\rm H_3^+ + e  \rightarrow &\ \rm H_2 + H  \tag{RA11}, \label{r:RA11}\\ 
\rm   \rightarrow &\  \rm H + H + H \tag{RA12}.
\label{r:RA12}
\end{align}
Thus, $\rm {{\it n}(H_3^+)}$ is in the form of:
\begin{align}
\rm {\it n}(H_3^+) =&\  \rm \frac{{\it n}(H_2)^2k_{RA5}k_{RA6}}{{\it n}(e)(k_{RA11}+k_{RA12})[{\it n}(HI)k_{RA3}+{\it n}(H_2)k_{RA6}]}.
\label{eq:h3p}
\end{align}
Note that equation \ref{eq:h3p} does not consider the destruction of H$^+_3$ by CO and O (see also equation 18 in \citet{Indriolo2012}), however, equation \ref{eq:h3p} is a good approximation in diffuse and translucent clouds since the destruction rate by electrons is more than two orders of magnitude higher than that of CO or O.

If we substitute equation \ref{eq:ohp1} with equation \ref{eq:hp2} and \ref{eq:h3p}, we have:
\begin{align}
&\ \rm {\it n}(OH^+){\it n}(H_2)k_{R4}  \notag \\
=&\  \rm {\it n}(O)\zeta_2[\frac{0.5(1.1-(0.98-0.88\xi){\it f}_{mol})k_{R1}{\it n}(H)}{{\it n}(e)k_{RA4}+{\it n}(O)k_{R1}} \notag \\
&\  \rm + 2.83 + \frac{0.22{\it f}_{mol}^2k_{RA6}k_{R3}{\it n}(H)}{{\it n}(e)(k_{RA11}+k_{RA12})} \times \notag \\
&\  \rm \frac{1}{(1-{\it f}_{mol})k_{RA3}+0.5{\it f}_{mol}k_{RA6}}] \notag \\
= &\ \rm {\it n}(O)\zeta_2\theta,
\label{eq:ohp2}
\end{align}
where $\theta$ is in the form of:
\begin{align}
\theta =&\  \rm \frac{0.5(1.1-(0.98-0.88\xi){\it f}_{mol})k_{R1}{\it n}(H)}{{\it n}(e)k_{RA4}+{\it n}(O)k_{R1}} \notag \\
&\  \rm + 2.83 + \frac{0.22{\it f}_{mol}^2k_{RA6}k_{R3}{\it n}(H)}{{\it n}(e)(k_{RA11}+k_{RA12})} \times \notag \\
&\  \rm \frac{1}{(1-{\it f}_{mol})k_{RA3}+0.5{\it f}_{mol}k_{RA6}},
\label{eq:theta}
\end{align}
which denotes the last term in equation \ref{eq:ohp2}. Combining equations \ref{eq:noh2} and \ref{eq:ohp2}, we thus obtain:
\begin{align}
\rm F[{\it n}(OH)] =&\  \rm {\it n}(O)\zeta_2\theta \times \notag \\
&\ \rm \frac{{\it n}(e)k_{R6}+{\it n}(H_2)k_{R5}(1-\delta)}{{\it n}(H_2)k_{R5}+{\it n}(e)(k_{R6}+k_{R7}+k_{R8})} \notag \\
=&\ \rm {\it n}(O)\zeta_2\theta \times \notag \\
&\ \rm [1-\frac{{\it n}(H_2)k_{R5}\delta+{\it n}(e)(k_{R7}+k_{R8})}{{\it n}(H_2)k_{R5}+{\it n}(e)(k_{R6}+k_{R7}+k_{R8})}].
\label{eq:noh3}
\end{align}

The destruction of OH is given by:
\begin{align}
\rm D[{\it n}(OH)] =&\  \rm {\it n}(OH)k_{pd}(R14) \notag \\
&\ \rm + {\it n}(OH){\it n}(C^+)(k_{R15}+k_{R16}) \notag \\
&\ \rm + {\it n}(OH){\it n}(H^+)k_{R17}.
\label{eq:doh1}
\end{align}

In equilibrium, the formation and destruction of OH reach a balance ($\rm F[{\it n}(OH)]$ = $\rm D[{\it n}(OH)]$).


\bibliography{reference}{}
\bibliographystyle{aasjournal}



\end{document}